\documentclass[utf8]{frontiersSCNS}

\usepackage{url,hyperref,lineno,microtype}
\usepackage[onehalfspacing]{setspace}
\usepackage{eurosym}
\usepackage{gensymb}
\usepackage{lscape}
\usepackage{rotating}
\usepackage{amsmath}
\usepackage{amsfonts}
\usepackage{sistyle}
\usepackage{float}
\usepackage{graphicx}
\usepackage{diagbox}
\usepackage{setspace}
\usepackage{longtable}

\def\keyFont{\fontsize{8}{11}\helveticabold }
\def\firstAuthorLast{Berger {et~al.}} %use et al only if is more than 1 author
\def\Authors{Mathias Berger\,$^{1,*}$, David Radu\,$^{1}$, Ghislain Detienne\,$^{2}$, Thierry Deschuyteneer\,$^{2}$, Aurore Richel\,$^{3}$ and Damien Ernst\,$^{1}$}
\def\Address{$^{1}$Department of Electrical Engineering and Computer Science, University of Li{\`e}ge, Li{\`e}ge, Belgium \\
$^{2}$ Fluxys SA, Brussels, Belgium\\
$^{3}$ Laboratory of Biomass and Green Technologies, Gembloux Agro-Bio Tech - University of Li{\`e}ge, Gembloux, Belgium}
\def\corrAuthor{Mathias Berger}

\def\corrEmail{mathias.berger@uliege.be}

\begin{document}
\onecolumn
\firstpage{1}

\title[Remote Renewable Hubs For Carbon-Neutral Synthetic Fuel Production]{Remote Renewable Hubs For Carbon-Neutral Synthetic Fuel Production} 

\author[\firstAuthorLast ]{\Authors} %This field will be automatically populated
\address{} %This field will be automatically populated
\correspondance{} %This field will be automatically populated

\extraAuth{}% If there are more than 1 corresponding author, comment this line and uncomment the next one.
%\extraAuth{corresponding Author2 \\ Laboratory X2, Institute X2, Department X2, Organization X2, Street X2, City X2 , State XX2 (only USA, Canada and Australia), Zip Code2, X2 Country X2, email2@uni2.edu}

\maketitle

\begin{abstract}
This paper studies the economics of carbon-neutral synthetic fuel production from renewable electricity in remote areas where high-quality renewable resources are abundant. To this end, a graph-based optimisation modelling framework directly applicable to the strategic planning of remote renewable energy supply chains is proposed. More precisely, a (hyper)graph abstraction of planning problems is introduced, wherein nodes can be viewed as optimisation subproblems with their own parameters, variables, constraints and local objective. Nodes typically represent a subsystem such as a technology, a plant or a process. Hyperedges, on the other hand, express the connectivity between subsystems. The framework is leveraged to study the economics of carbon-neutral synthetic methane production from solar and wind energy in North Africa and its delivery to Northwestern European markets. The full supply chain is modelled in an integrated fashion, which makes it possible to accurately capture the interaction between various technologies on hourly time scales. Results suggest that the cost of synthetic methane production and delivery would be slightly under 150 \euro/MWh (higher heating value) by 2030 for a system supplying 10 TWh annually and relying on a combination of solar photovoltaic and wind power plants, assuming a uniform weighted average cost of capital of 7\%. A comprehensive sensitivity analysis is also carried out in order to assess the impact of various techno-economic parameters and assumptions on synthetic methane cost, including the availability of wind power plants, the investment costs of electrolysis, methanation and direct air capture plants, their operational flexibility, the energy consumption of direct air capture plants, and financing costs. The most expensive configuration (around 200 \euro/MWh) relies on solar photovoltaic power plants alone, while the cheapest configuration (around 88 \euro/MWh) makes use of a combination of solar PV and wind power plants and is obtained when financing costs are set to zero. 
\tiny
 \keyFont{ \section{Keywords:} optimisation, renewable energy, carbon neutral, synthetic fuels, remote supply chain, linear programming, structured models, graph}
\end{abstract}

\section{Introduction}

Electricity generation from renewable resources combined with wide-ranging electrification has been a mainstay of European climate and energy policies, with the primary goal of decarbonising the power sector as well as other carbon-intensive sectors. 

Major obstacles to such endeavours have nevertheless surfaced in recent years. Firstly, sectors like aviation, shipping, heating or industry have proved difficult to fully electrify. Indeed, feedstocks and energy carriers with specific properties such as a high energy density are typically required \citep{Eveloy2021}. Hence, the production of carbon-neutral synthetic fuels and feedstocks from renewable electricity has been the focus of a growing body of literature. For example, the synthesis of carbon-neutral hydrogen \citep{Borgschulte2016}, methane \citep{Biswas2020}, methanol \citep{Centi2020} and ammonia \citep{Ghavam2021} have all been considered. A number of demonstration projects have been carried out as well \citep{Wulf2020}. Secondly, it has become clear that the technical renewable potential of some European countries (i.e., the maximum amount of renewable electricity that may produced within a country's borders and exclusive economic zone, while accounting for a variety of land eligibility constraints \citep{Ryberg2018}) is insufficient to supply current energy demand levels (e.g., in densely-populated countries like Belgium \citep{Berger2020,Limpens2020} or the United Kingdom \citep{MacKay2008}). It is still unclear whether pooling renewable resources at the European level would alleviate the problem. On the other hand, it is well-documented that social acceptance issues tend to compound it \citep{Segreto2020}.

A simple solution consists in harvesting renewable resources in remote areas where they are abundant, synthesising carbon-neutral fuels or feedstocks using renewable electricity and transporting them back to demand centres \citep{Fasihi2015,Chapman2017,Heuser2019}. However, two conditions must be satisfied for such an approach to be worth pursuing. Firstly, transport should be energy-efficient and cost-effective. This will often depend on the physics of the commodity considered and the maturity of technologies available to handle it. Secondly, very-high-quality renewable resources should be tapped. The quality of such resources is typically estimated via the annual capacity factor of a given technology harnessing them, which directly reflects the amount of electricity that may be produced per unit capacity. Since renewable power generation technologies usually have very low operating costs, the higher the capacity factor, the lower the electricity cost. Regions with outstanding resources and vast technical potential include Patagonia (wind) \citep{Heuser2019}, North Africa (sun and wind) \citep{Fasihi2015} and Greenland (wind) \citep{Radu2019}. Providing an accurate quantitative assessment of the economics and efficiency of such remote renewable energy supply chains and pathways is critical to evaluate future sustainable energy supply options available to policy makers and society at large as well as to identify where to direct future research and innovation efforts.

From a conceptual standpoint, a supply chain can be viewed as a networked system composed of dynamical subsystems interacting with each other. In order to tackle the problem formulated above, the collection of processes and technologies forming a remote renewable energy supply chain must be analysed in an integrated fashion, which makes it possible to properly capture the interactions between subsystems in space and time. In addition, a sufficient level of technical detail and temporal resolution should be used to properly model their operation \citep{Poncelet2016}. This paper formalises these considerations and proposes a graph-based optimisation modelling framework directly applicable to the strategic planning and analysis of remote renewable energy supply chains. More precisely, a graph abstraction of planning problems is introduced, wherein nodes can be viewed as optimisation subproblems with their own parameters, variables, constraints and local objective, and typically represent a subsystem such as a technology, a plant or a process. Edges, on the other hand, express the connectivity between subsystems. The framework is then leveraged to study the economics of carbon-neutral synthetic methane production from renewable electricity and atmospheric carbon dioxide in North Africa and its delivery to Northwestern European markets. Synthetic methane is an appealing carbon-neutral energy carrier, as some downstream transport infrastructure is readily available in Northwestern European countries, and the liquefied methane chain is mature and cost-effective \citep{lngcarrierefficiency}. It would also obviate the need for replacing or upgrading appliances and processes presently used for residential heating and in industry that a switch to other fuels would entail. In this paper, the carbon-neutral synthetic methane supply chain is modelled end-to-end, from power generation in North Africa to methane regasification in Northwestern Europe. A detailed description of each process and technology is provided, along with comprehensive data resources. The modelling framework also served as a basis for the development of an open source optimisation modelling language \citep{GBOMLtutorial} and tool \citep{GBOML}. In the interest of transparency, the input files and full data enabling others to reproduce the analyses presented in this paper are also available in the associated repository \citep{GBOML}.

This paper is structured as follows. Section \ref{related_work} reviews the relevant literature. Section \ref{methodology} details the proposed modelling framework, while Sections \ref{casestudy} and \ref{results} describe the case study and discuss results, respectively. Finally, Section \ref{conclusion} concludes the paper and discusses future work directions.

\section{Literature Review}\label{related_work}

To the best of the authors' knowledge, \cite{Hashimoto1999} were the first to suggest the production of hydrogen from renewable electricity in remote areas followed by the synthesis of hydrocarbons using captured carbon dioxide as a means of producing carbon-neutral fuels. The paper, however, did not provide a quantitative techno-economic analysis of the proposed supply chains. By contrast, \cite{Zeman2008} performed one of the first quantitative economic analyses of carbon-neutral synthetic fuel production using carbon-neutral hydrogen and atmospheric carbon dioxide. Production cost estimates for this route were found to be between 23.5 and 30.0 US\$/GJ (which would roughly correspond to 74.1 and 94.6 \euro/MWh, using the 2020 average exchange rate of \$1.142 for 1.0\euro). The production of carbon-neutral synthetic methane and liquid fuels in remote areas with abundant renewable resources has been considered in \cite{Fasihi2015} and \cite{Fasihi2017}, respectively. In the first study, the authors estimate that the cost of producing synthetic methane from renewable electricity in the Maghreb and North Africa (specifically in central and southern Algeria) and delivering it to Japan could be around 65-75 \euro/MWh by 2030 for a hybrid solar-wind system, assuming a uniform weighted average cost of capital (WACC) of 7\%. It is not specified whether the higher heating value (HHV) or the lower heating value (LHV) of methane was used to compute these costs. In the second study, the cost of producing synthetic methane in the same region and delivering it to Finland is found to be between 100-110 \euro/MWh (HHV) by 2030 and between 90-100 \euro/MWh (HHV) by 2040, respectively, using a WACC of 7\%. Finally, the economics of carbon-neutral fuel production is also analysed in \cite{solarpvcosts}. Cost estimates close to 140-150 \euro/MWh (LHV) by 2030 and 110 \euro/MWh (LHV) by 2050 (using a WACC of 6\% in both cases) are found for synthetic methane production in North Africa (specifically in central and southern Algeria) and delivery to Germany based on both solar energy alone and hybrid systems combining solar and wind power plants. 

It is also informative to review the modelling approaches followed in these studies. Firstly, \cite{Zeman2008} do not specify the technologies used to implement the various conversion processes, and instead rely on a set of assumptions about conversion efficiencies and the cost of producing input commodities (in stoichiometric proportions) to come up with a cost estimate for the final product. Then, \cite{Fasihi2015} resort to a so-called \textit{annual-basis model} estimating the annual number of equivalent full load hours of renewable power production in order to calculate electricity and synthetic methane costs based on a set of techno-economic assumptions. This method is equivalent to estimating annual power production and costs using an average capacity factor value, and the model is therefore not temporally-resolved. A so-called \textit{hourly-basis model} enabling the sizing of solar photovoltaic (PV) and wind power plants is mentioned in \cite{Fasihi2015,Fasihi2017}, but no mathematical model is explicitly described and no computer code implementing it is made available, which makes the approach difficult to interpret and scrutinise. Somewhat surprisingly, very minor differences in cost estimates are observed between the annual-basis and hourly-basis models in \cite{Fasihi2015}. In \cite{solarpvcosts}, an annual full load hour model similar to that of \cite{Fasihi2015} is used. For systems driven by variable renewable energy resources, it has been shown that using a high temporal resolution (e.g., hourly) and adopting a proper level of technical detail (i.e., representing the flexibility of technologies, or lack thereof) is key for accurately sizing plants and estimating both investment and operating costs properly \citep{Poncelet2016}. It is worth noting that the aforementioned papers rely on models that have both a very low level of technical detail and a very low temporal resolution. Furthermore, none of these models makes it possible to design the supply chain in an integrated fashion while properly accounting for interactions between subsystems. Similar shortcomings can be found in studies focussing on other energy carriers such as hydrogen \citep{Dagdougui2012,Heuser2019}.

The design and analysis of energy systems and supply chains has often been tackled using mathematical programming techniques in the literature \citep{Garcia2015,Conejo2016}. Different classes of models may be used, ranging from linear and mixed-integer linear programs (LPs and MILPs) to nonlinear and mixed-integer (possibly nonconvex) nonlinear programs (NLPs and MINLPs) \citep{Biegler2004}. Parameter uncertainty may also be taken into account \citep{Sahinidis2004}. The type of model used typically depends on the research scope, the available computational resources and the data at hand. For example, the design of a single piece of equipment used in a process may require NLP or MINLP models to accurately represent its physics and operating modes \citep{Grossmann2002}. On the other hand, supply chains can be viewed as collections of interconnected plants or processes, which themselves rely on a variety of complex pieces of equipment. Representing each of them in their full complexity would require vast amounts of data and result in intractable models. Thus, for the purpose of strategic or high-level system design analyses, aggregate plant models are typically employed \citep{Chen2017,Montastruc2019}. In such models, mass and energy conservation laws are enforced at plant level while accounting for basic operational constraints. Mass and energy balances are also enforced between interconnected plants in order to guarantee consistency of flows at system level. Such approaches, which usually rely on LP or MILP models, have for instance been applied to the design of integrated biorefineries \citep{Kokossis2015}, the design of power-to-syngas processes \citep{Maggi2020} and power-to-chemicals networks \citep{Schack2016,Liesche2019}. Such an approach is adopted in this paper, as discussed next.

\section{Methodology}\label{methodology}

This section formally introduces the abstract graph-based optimisation modelling framework and describes how it can be applied in the context of strategic energy supply chain planning.

\subsection{Graph-Based Optimisation Modelling Framework}\label{modellingframework}

In this paper, supply chain planning problems are formulated as structured linear programs. These problems typically involve the optimisation of discrete-time dynamical systems over a finite time horizon and exhibit a natural block structure that may be encoded by a sparse graph or hypergraph \citep{Gallo1993}. A graph abstraction is therefore employed to represent them, wherein nodes model optimisation subproblems, while hyperedges express the relationships between nodes. A global discretised time horizon and associated set of time periods common to all nodes are also defined. Each node is equipped with a set of so-called \textit{internal} and \textit{external} (or \textit{coupling}) variables. A set of constraints is also defined for each node, along with a local objective function representing its contribution to a system-wide objective. Finally, for each hyperedge, constraints involving the coupling variables of the nodes to which it is incident are defined in order to express the relationships between nodes. In the following paragraphs, we formally define variables, constraints, objectives and formulate the abstract model that encapsulates the class of problems considered.

Let $T$ be the time horizon considered, let $\mathcal{T} = \{0, 1, \dots, T - 1\}$ be the associated set of time periods, and let $\mathcal{G} = (\mathcal{N}, \mathcal{E})$ be a (possibly directed) hypergraph encoding the block structure of the problem considered, with node set $\mathcal{N}$ and hyperedge set $\mathcal{E} \subseteq 2^\mathcal{N}$ (i.e., each hyperedge corresponds to a subset of nodes). Let $X^n \in \mathcal{X}^n$ and $Z^n \in \mathcal{Z}^n$ denote the collection of internal and coupling variables defined at node $n \in \mathcal{N}$. Note that all variables are assumed to take values in continuous sets (i.e., $\mathcal{X}^n$ and $\mathcal{Z}^n$ are continuous). In addition, for any hyperedge $e \in \mathcal{E}$, let $Z^e = \{Z^n | n \in e\}$ denote the collection of coupling variables associated with each node to which this hyperedge is incident.

Let $F^n$ denote the function defining the local objective at node $n \in \mathcal{N}$. In this paper, we consider scalar objectives of the form
\begin{equation}
F^n(X^n, Z^n) = f_0^n(X^n, Z^n) + \sum_{t \in \mathcal{T}} f^n(X^n, Z^n, t),
\end{equation}
where $f_0^n$ and $f^n$ are (scalar) affine functions of $X^n$ and $Z^n$.

Both equality and inequality constraints may be defined at each node $n \in \mathcal{N}$. More precisely, an arbitrary number of constraints that can each be expanded over a subset of time periods may be defined. Hence, we consider equality constraints of the form\vspace{-5pt}
\begin{equation}
\label{eq:model:eq}
h_k^n(X^n, Z^n, t) = 0, \mbox{ } \forall t \in \mathcal{T}_k^n,\vspace{-5pt}
\end{equation}
with (scalar) affine functions $h_k^n$ and index sets $\mathcal{T}_k^n \subseteq \mathcal{T}$, $k = 1, \ldots, K^n$, as well as inequality constraints
\begin{equation}
\label{eq:model:ineq}
g_k^n(X^n, Z^n, t) \le 0, \mbox{ } \forall t \in \bar{\mathcal{T}}_k^n,\vspace{-5pt}
\end{equation}
with (scalar) affine functions $g_k^n$ and index sets $\bar{\mathcal{T}}_k^n \subseteq \mathcal{T}$, $k = 1, \ldots, \bar{K}^n$.

Likewise, both equality and inequality constraints may be defined over any hyperedge $e \in \mathcal{E}$. These constraints, however, can only involve the coupling variables of the nodes to which hyperedge $e \in \mathcal{E}$ is incident (i.e., nodes such that $n \in e$). More precisely, let $H^e$ and $G^e$ be affine functions of $Z^e$ used to define the equality and inequality constraints associated with a given hyperedge $e\in\mathcal{E}$.

Using this notation, the class of problems that can be represented in this framework reads
\begin{equation}
\label{eq:model:prob}
\begin{array}{rl}
\min & \sum_{n \in \mathcal{N}} F^n(X^n, Z^n) \vspace{3pt}\\
\hbox{s.t.} & h_k^n(X^n, Z^n, t) = 0, \mbox{ } \forall t \in \mathcal{T}_k^n, \hspace{2pt} k = 1, \ldots K^n, \hspace{2pt} \forall n \in \mathcal{N} \vspace{3pt}\\
& g_k^n(X^n, Z^n, t) \le 0, \mbox{ } \forall t \in \bar{\mathcal{T}}_k^n, \hspace{2pt} k = 1, \ldots \bar{K}^n, \hspace{2pt} \forall n \in \mathcal{N} \vspace{3pt}\\
& H^e(Z^e) = 0, \hspace{2pt} \forall e \in \mathcal{E} \vspace{3pt}\\
& G^e(Z^e) \le 0, \hspace{2pt} \forall e \in \mathcal{E} \vspace{3pt}\\
& X^n \in \mathcal{X}^n, Z^n \in \mathcal{Z}^n, \hspace{2pt} \forall n \in \mathcal{N}.
\end{array}
\end{equation}

\begin{figure*}[h!]
\centering
\includegraphics[scale=0.5]{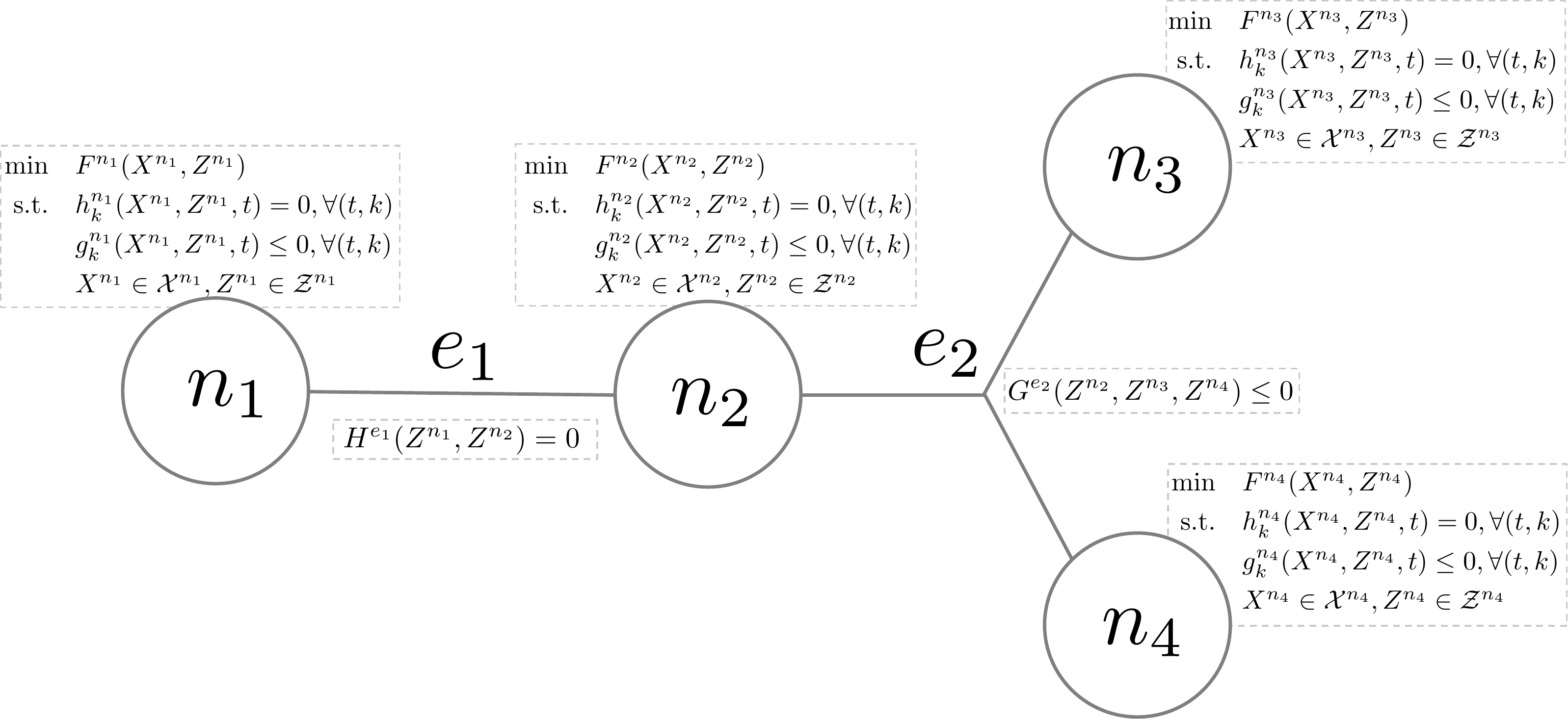}
\caption{Graph abstraction of a hypothetical problem whose block structure is represented by four nodes (i.e., $\mathcal{N} = \{n_1, n_2, n_3, n_4\}$) and two hyperedges (i.e., $\mathcal{E} = \{e_1, e_2\}$, with $e_1 = \{n_1,  n_2\}$ and $e_2 = \{n_2, n_3, n_4\}$). Note that $e_1$ has only equality constraints while $e_2$ has only inequality constraints.}
\label{hypergraph_schematic}
\end{figure*}

Figure \ref{hypergraph_schematic} schematically illustrates the class of problems that can be modelled in this framework.

\subsection{Application to Energy Supply Chains}\label{applicationenergysystems}

The framework presented in Section \ref{modellingframework} can be readily leveraged to model energy systems and supply chains. In this case, nodes typically represent a technology, a plant or a process, while hyperedges may be used to enforce some coupling between plants. Introducing a few generic (parametrised) nodes and hyperedges often suffices to model a broad range of system configurations. In the following, some key modelling assumptions are introduced, along with two generic nodes and one generic hyperedge, namely \textit{conversion} and \textit{storage} nodes, as well as \textit{conservation} hyperedges.

\subsubsection{Modelling Assumptions}\label{modellingassumptions}

\textbf{Central Planning \& Operation} Investment decisions are made by a single entity that also operates the system, and whose goal is to minimise total system costs.

\textbf{Perfect Foresight \& Knowledge} The entity planning and operating the system has perfect foresight and knowledge, that is, future weather events and demand patterns, as well as all technical and economic parameters are assumed to be known with certainty.

\textbf{Investment \& Operational Decisions} A static investment model is used, whereby investment decisions are made at the beginning of the time horizon and assets are immediately available. Operational decisions are made at hourly time steps. The investment and operational problems are solved simultaneously.

\textbf{Technology \& Process Models} The sizing and operation of technologies are modelled via a set of affine input-output relations that typically express mass and energy balances at plant or process level. Input or output dynamics are considered for some technologies, but only storage technologies have a simple state space representation.

\subsubsection{Nodes}

\textbf{Preliminaries} In the following developments, Latin letters denote optimisation variables and indices, while Greek letters indicate parameters.

\textbf{Conversion} Let $n \in \mathcal{N}$ be a node representing a so-called \textit{conversion technology} that processes a set of commodities (e.g., an electrolysis plant splits water into hydrogen and oxygen using an electric current and therefore processes four commodities). Commodity flows are modelled as external variables. An index $i \in \mathcal{I}^n$ is thus assigned to each commodity (i.e., each $i \in \mathcal{I}^n$ corresponds to a time-indexed external variable). The processing of commodities by technology $n$ is modelled via a set of linear equations linking the flow of a reference commodity $r \in \mathcal{I}^n$ (e.g., hydrogen may be taken as the reference commodity for electrolysis plants) to the flow of all other commodities $i \in \mathcal{I}^n\setminus\{r\}$, which read 
\begin{equation}
q_{rt}^n - \phi_{i}^n q_{i(t+\tau_{i}^n)}^n = 0, \mbox{ } \forall i \in \mathcal{I}^n\setminus\{r\}, \mbox{ } \forall t \in \mathcal{T}^n,
\label{conversion}
\end{equation}
where $q_{it}^n \in \mathbb{R}_+$ represents the flow of commodity $i$ at time $t$, $\phi_{i}^n \in \mathbb{R}_+$ is the so-called \textit{conversion factor} between commodity $r$ and $i$ (which may be derived, e.g., from stoichiometric coefficients or the enthalpy of the underlying reaction), while $\tau_{i}^n \in \mathbb{N}$ is the amount of time that may be required for the conversion process to take place and $\mathcal{T}^n \subseteq \mathcal{T}$ is a suitable subset of time periods. The capacity of a technology is typically modelled as an internal variable and defined as the maximum flow of a reference commodity $r' \in \mathcal{I}^n$ according to which the technology is sized. Note that $r'$ may be different from $r$ (e.g., the size of electrolysis plants is typically expressed in terms of their electrical capacity, although hydrogen may be the reference commodity used in Equation \eqref{conversion}). Since a static investment model is considered, capacity deployments occur at the beginning of the time horizon and remain constant throughout, i.e., 
\begin{equation}
K_0^n - K_t^n = 0, \mbox{ } \forall t \in \mathcal{T}\setminus\{0\},
\label{conversioncapacitystatic}
\end{equation}
where $K_t^n \in \mathbb{R}_+$ denotes the new capacity of technology $n$. In the following, $K^n$ will be used as shorthand for $K_0^n$. Thus, the total capacity of technology $n$ is defined via
\begin{equation}
q_{r't}^n - \pi_{t}^n (\underbar{$\kappa$}^{n} + K^{n}) \le 0, \mbox{ } \forall t \in \mathcal{T},
\label{sizing}
\end{equation}
where $\pi_{t}^n \in [0, 1]$ indicates the availability of technology $n$ at time $t$ and $\underbar{$\kappa$}^n \in \mathbb{R}_+$ represents the existing capacity. The so-called \textit{availability parameter} $\pi_{t}^n$ may for instance represent the instantaneous capacity factor of a renewable power plant. The maximum capacity of a technology may be bounded, which leads to the introduction of an additional constraint,
\begin{equation}
(\underbar{$\kappa$}^{n} + K^{n}) - \bar{\kappa}^{n} \le 0,
\label{potential}
\end{equation}
with $\bar{\kappa}^n \in \mathbb{R}_+$ the maximum capacity of technology $n$ that may be installed. A variety of operational constraints may also be considered. For instance, some conversion technologies may have a limited operating range, and may only work if a minimum flow of commodity $i \in \mathcal{I}^n$ is maintained, which can be expressed as
\begin{equation}
\mu^{n} (\underbar{$\kappa$}^{n} + K^{n}) - \frac{\phi_{i}^n}{\phi_{r'}^n}q_{it}^n \le 0, \mbox{ } \forall t \in \mathcal{T},
\label{mustrun}
\end{equation}
where $\mu^n \in [0, 1]$ represents the minimum operating level (as a fraction of the installed capacity). Since the technology is sized with respect to the flow of commodity $r'$, the flow of a commodity $i \ne r'$ must be scaled by the ratio of conversion factors in Eq. \eqref{mustrun}. The rate at which the flow of commodity $i \in \mathcal{I}^n$ can vary may also be limited, leading to the introduction of so-called ramping constraints,
\begin{equation}
\frac{\phi_{i}^n}{\phi_{r'}^n}(q_{it}^n - q_{i(t-1)}^n) - \Delta_{i,+}^{n} (\underbar{$\kappa$}^{n} + K^{n}) \le 0, \mbox{ } \forall t \in \mathcal{T}\setminus\{0\},
\label{rampup}
\end{equation}
and
\begin{equation}
\frac{\phi_{i}^n}{\phi_{r'}^n}(q_{i(t-1)}^n - q_{it}^n) - \Delta_{i,-}^{n} (\underbar{$\kappa$}^{n} + K^{n}) \le 0, \mbox{ } \forall t \in \mathcal{T}\setminus\{0\},
\label{rampdown}
\end{equation}
with $\Delta_{i,+}^{n} \in [0,1]$ and $\Delta_{i,-}^{n} \in [0,1]$ the maximum rates at which flows can be ramped up and down (as a fraction of the installed capacity per unit time), respectively. Finally, the local objective function associated with this node reads
\begin{equation}
F_n = \nu (\zeta^{n} + \theta_f^n) K^{n} + \sum_{t \in \mathcal{T}} \theta_{t,v}^n q_{r't}^n \delta t,\label{objectiveconversion}
\end{equation}
where $\nu \in \mathbb{N}$ is the number of years spanned by the optimisation horizon, $\zeta^n \in \mathbb{R}_+$ represents the (annualised) investment cost (also known as capital expenditure, CAPEX), $\theta_f^n \in \mathbb{R}_+$ models fixed operation and maintenance (FOM) costs and $\theta_{t,v}^n \in \mathbb{R}_+$ represents variable operation and maintenance (VOM) costs, which may be time-dependent.

\textbf{Storage} Let $n \in \mathcal{N}$ be a node representing a \textit{storage technology}. A storage technology is assumed to hold one commodity, although its operation may involve other commodities (e.g., a compressed hydrogen storage system stores hydrogen but requires electricity to drive compressors). The inventory level of the storage system is defined as an internal variable, while the charge and discharge flows are defined as external variables, respectively. Let $i \in \mathcal{I}^n$ and  $j \in \mathcal{I}^n$ be the indices of the in/outflows of the commodity stored in technology $n$, respectively. Then, the basic equation governing the operation of storage systems describes the inventory level dynamics and reads
\begin{equation}
e_{t+1}^n - (1-\eta_{S}^n) e_{t}^n - \eta_{+}^{n} q_{it}^n + \frac{1}{\eta_{-}^{n}} q_{jt}^n = 0, \mbox{ } \forall t \in \mathcal{T} \setminus\{T-1\},
\label{storagedynamics}
\end{equation}
where $e_{t}^n \in \mathbb{R}_+$ is the inventory level at time $t$, $q_{i_ut}^n \in \mathbb{R}_+$ and $q_{i_yt}^n \in \mathbb{R}_+$ represent commodity in- and outflows at time $t$, respectively, $\eta_S^n \in [0, 1]$ is the self-discharge rate, $\eta_+^n \in [0, 1]$ is the charge efficiency and $\eta_-^n \in [0, 1]$ is the discharge efficiency. Charging a storage system may also require the consumption of another commodity $l \in \mathcal{I}^n, l \ne i, j$ (e.g., electricity consumed by compressors), which is typically modelled via an additional external variable $q_{lt}^n \in \mathbb{R}_+$ and equations
\begin{equation}
q_{lt}^n - \phi_{i}^n q_{it}^n = 0, \mbox{ } \forall t \in \mathcal{T}.
\label{conversionstorage}
\end{equation}
In order to avoid spurious transient effects in storage operation, inventory levels are typically required to be equal at the beginning and at the end of the optimisation horizon,
\begin{equation}
e_{0}^n - e_{T-1}^n = 0.
\label{storagecyclicity}
\end{equation}
The stock capacity of the storage technology is modelled as an internal variable and it is defined by the maximum inventory level.   Since a static investment model is used, the stock capacity is constant throughout the entire time horizon, i.e.,
\begin{equation}
E_0^n - E_t^n = 0, \mbox{ } \forall t \in \mathcal{T}\setminus\{0\},
\label{storagestockstatic}
\end{equation}
where $E_t^n \in \mathbb{R}_+$ is the new capacity. In the following, $E^n$ will be used as shorthand for $E_0^n$. The total storage capacity is therefore defined via
\begin{equation}
e_{t}^n - (\underbar{$\epsilon$}^{n} + E^{n}) \le 0, \mbox{ } \forall t \in \mathcal{T},
\label{storagestocksizing}
\end{equation}
where $\underbar{$\epsilon$}^n \in \mathbb{R}_+$ denotes the existing stock capacity. Note that the total stock capacity itself may be constrained,
\begin{equation}
(\underbar{$\epsilon$}^{n} + E^{n}) - \bar{\epsilon}^{n} \le 0,
\label{storagepotential}
\end{equation}
with $\bar{\epsilon}^n \in \mathbb{R}_+$ the maximum stock capacity that may be deployed. In addition, some storage technologies may require a minimum inventory level to be maintained, which can be expressed as
\begin{equation}
\sigma^{n}(\underbar{$\epsilon$}^{n} + E^{n}) - e_{t}^n \le 0, \mbox{ } t \in \mathcal{T},
\label{storageminimumsoc}
\end{equation}
where $\sigma^n \in [0, 1]$ represents the minimum inventory level (as a fraction of the stock capacity). The maximum inflow capacity is sized independently of the stock capacity. For example, in the case of battery storage systems, this implies that the energy-to-power ratio is not fixed \textit{a priori}. The maximum inflow is modelled using an internal variable that is also constant throughout the time horizon considered, as in Eq. \eqref{conversioncapacitystatic}. It is defined as follows
\begin{equation}
q_{it}^n - (\underbar{$\kappa$}^{n} + K^{n}) \le 0, \mbox{ } \forall t \in \mathcal{T},
\label{storageflowplussizing}
\end{equation}
where $\underbar{$\kappa$}^n \in \mathbb{R}_+$ denotes the existing flow capacity and $K^n \in \mathbb{R}_+$ is used as shorthand for the new capacity. The maximum in- and outflows may be asymmetric, depending on the properties of the underlying technology, which is modelled via
\begin{equation}
q_{jt}^n - \rho^{n} (\underbar{$\kappa$}^{n} + K^{n}) \le 0, \mbox{ } \forall t \in \mathcal{T},
\label{storageflowminussizing}
\end{equation}
where $\rho^n \in \mathbb{R}_+$ represents the maximum discharge-to-charge ratio. Finally, the local objective function associated with this node reads
\begin{equation}
F^n = \Big[\nu (\varsigma^{n} + \vartheta^{n}_{f}) E^{n} + \sum_{t \in \mathcal{T}} \vartheta_{t,v}^n e_{t}^n\Big] + \Big[\nu (\zeta^{n} + \theta^n_f) K^{n} + \sum_{t \in \mathcal{T}} \theta_{t,v}^n q_{it}^n \delta t \Big].
\label{objectivestorage}
\end{equation}
where $\varsigma^n \in \mathbb{R}_+$ and $\zeta^n \in \mathbb{R}_+$ represent the stock and flow components of CAPEX, $\vartheta_f^n \in \mathbb{R}_+$ and $\theta_f^n \in \mathbb{R}_+$ model the stock and flow components of FOM costs, while $\vartheta_{t,v}^n \in \mathbb{R}_+$ and $\theta_{t,v}^n \in \mathbb{R}_+$ represent the stock and flow components of VOM costs, which may be time-dependent.

\subsubsection{Hyperedges}

\textbf{Conservation} Let $e \in \mathcal{E}$ be a so-called \textit{conservation hyperedge} that enforces local flow conservation of some commodity. More precisely, one commodity is associated with a given conservation hyperedge. Let $i \in \cap_{n \in e} \mathcal{I}^n$ denote this commodity (i.e., each node $n \in e$ has an external variable representing a flow of commodity $i$). In addition, let us assume that hyperedge $e$ is directed (i.e., it can be partitioned into two disjoint subsets $e_T$ and $e_H$ that are called its tail and head, respectively). Roughly speaking, $e$ can be interpreted as "going from nodes in $e_T$ to nodes in $e_H$". Then, flow conservation of commodity $i$ over hyperedge $e$ can simply be expressed as
\begin{equation}
\sum_{n \in e_T} q_{it}^n - \sum_{n \in e_H} q_{it}^n - \lambda_{t}^e = 0, \mbox{ } \forall t \in \mathcal{T},
\label{conservationhyperedge}
\end{equation}
where the first two sums on the left-hand side represent the aggregate flows from the nodes in $e_T$ and $e_H$ (whose signs depend on the orientation of $e$), while $\lambda_{t}^e \in \mathbb{R}$ represents exogenous withdrawals or injections that may take place over $e$ at each time period $t$. Note that Eq. \eqref{conservationhyperedge} may sometimes be relaxed to a greater-than-or-equal-to inequality constraint (in which case the net flow must exceed the injections/withdrawals at each time period).

\subsection{Implementation}
The graph-based modelling framework discussed in Section \ref{modellingframework} has been used as a basis for developing an optimisation modelling language for structured linear and mixed-integer linear programs called the graph-based optimisation modelling language (GBOML) \citep{GBOMLtutorial}. The language blends elements of both algebraic \citep{Kallrath2012} and object-oriented \citep{Schichl2004} modelling languages in order to facilitate problem encoding and post-processing, promote model re-use and improve portability. The full description of GBOML, which is beyond the scope of this paper, is detailed in a separate tutorial paper \citep{GBOMLtutorial}. A parser for GBOML, called the GBOML compiler, has also been implemented in Python 3.8 (using the PLY library), and is released as open source software \citep{GBOML}. The GBOML compiler directly interfaces with both commercial and open source LP and MILP solvers (namely Gurobi, CPLEX and Clp/Cbc), enabling users to model problems, interact with solver APIs, query solutions and retrieve post-processed results in an integrated fashion. For the sake of transparency and reproducibility, the input file that encodes the model and full data allowing one to reproduce the case study and results discussed in Sections \ref{casestudy} and \ref{results} are also provided in the GBOML repository \citep{GBOML}. One instance of the resulting linear programming model can be solved in about 10 minutes with the homogenous barrier algorithm (cross-over disabled) of Gurobi 9.1.1 on a laptop with 16 GB of RAM and a Quad-Core Intel i7 processor clocking at 2.6 GHz.

\section{Case Study} \label{casestudy}
\begin{figure*}[h!]
\centering
\includegraphics[scale=0.85]{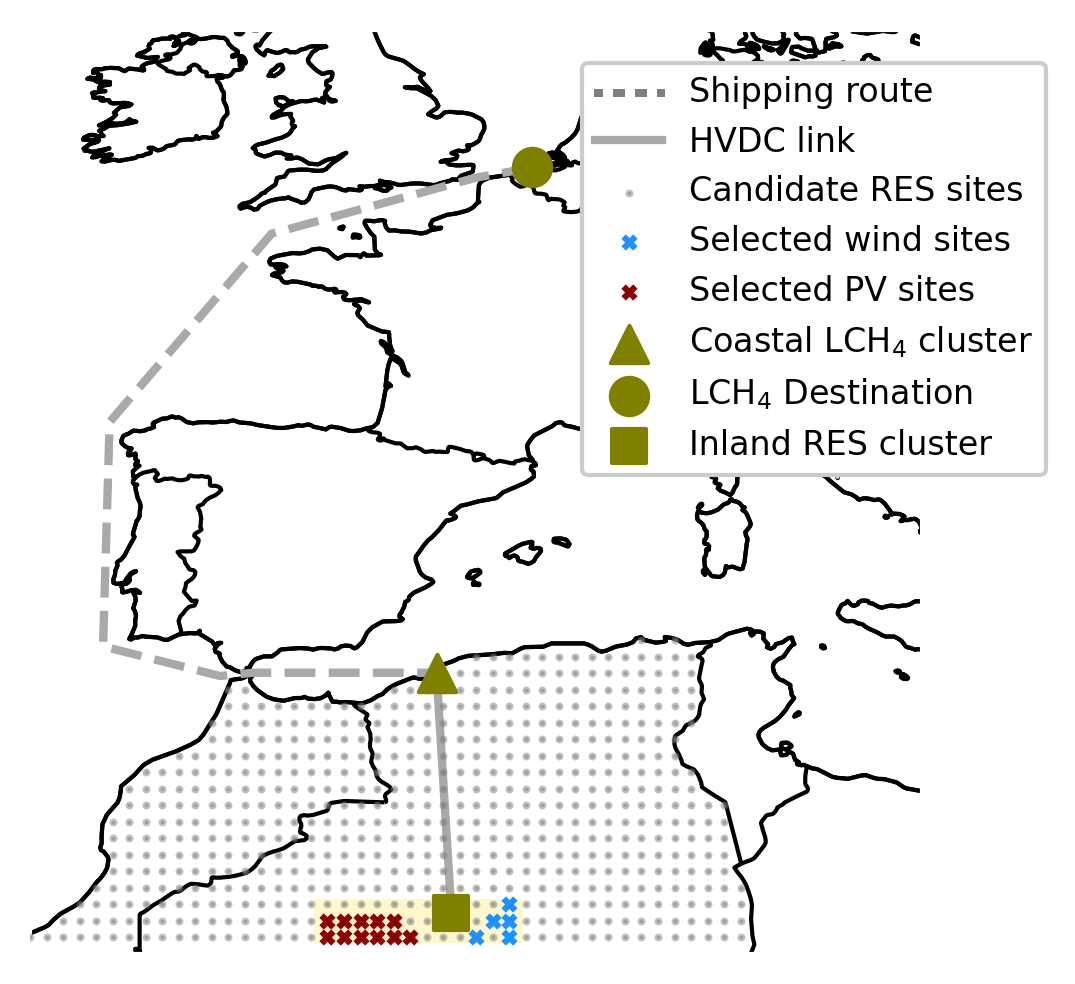}
\caption{Remote carbon-neutral methane supply chain: electricity is produced in a remote inland cluster in central Algeria and transported to a coastal cluster where carbon-neutral methane is synthesised and liquefied for export to Northwestern European markets.}
\label{map_schematic}
\end{figure*}

This case study aims to analyse the economics of producing carbon-neutral methane from renewable electricity in areas of North Africa where abundant and high-quality renewable resources are readily available, and exporting it to Northwestern European markets. More specifically, the entire supply chain is modelled and optimised in an integrated fashion over a time horizon of five years with hourly resolution (i.e., $T = 43824$, since 2016 is a leap year), from the remote generation of electricity to the synthesis and liquefaction of carbon-neutral methane in North Africa, to its eventual delivery and regasification at a Northwestern European gas terminal. Figure \ref{map_schematic} schematically displays the supply chain considered. 

\begin{figure*}[h!]
\centering
\includegraphics[scale=0.75]{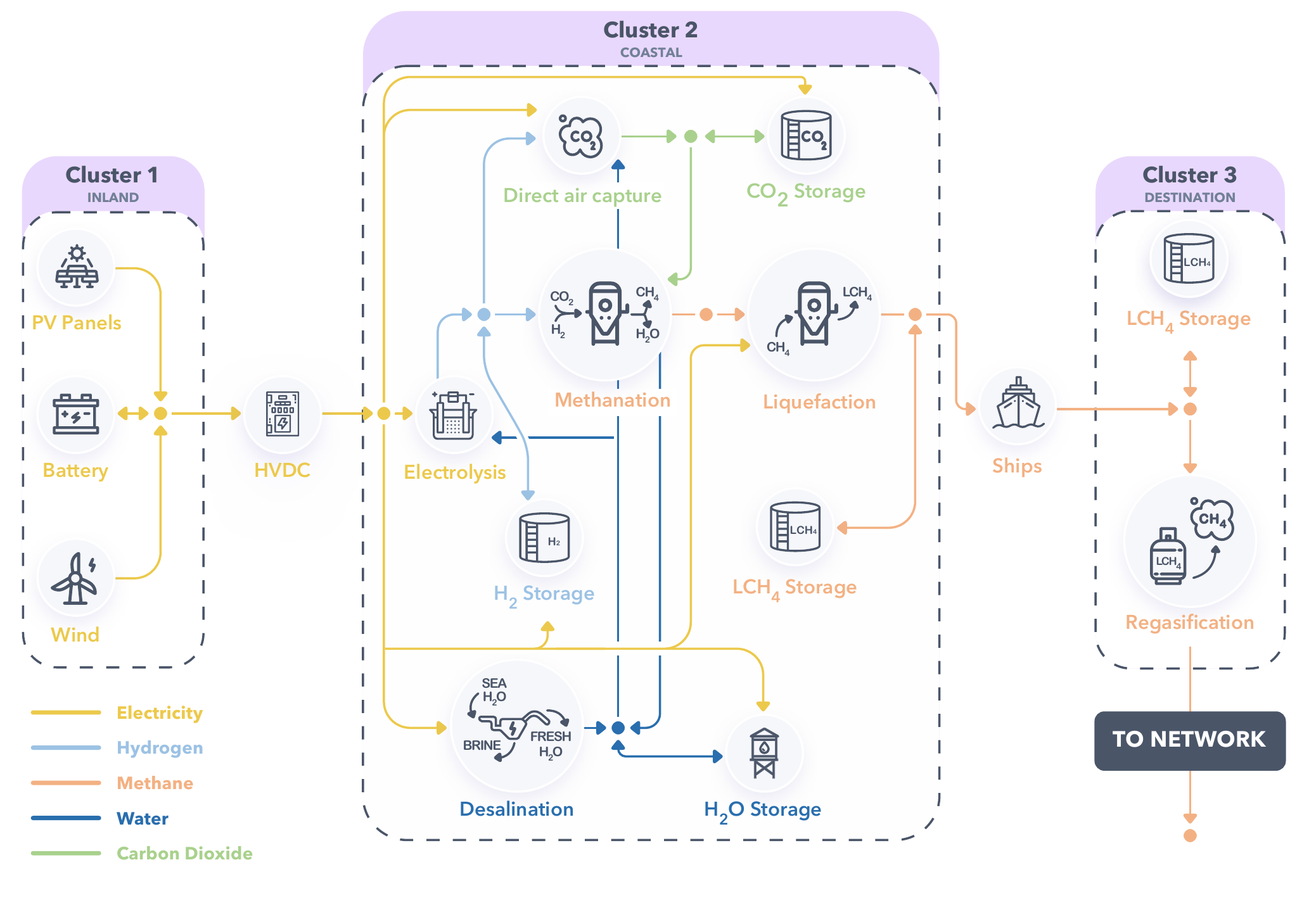}
\caption{Remote hub system configuration. Icons represent conversion or storage nodes, while bullets and arrows schematically represent conservation hyperedges.}
\label{systemconfiguration}
\end{figure*}

\subsection{System Configuration}
A more detailed representation of the system configuration considered in this study is shown in Figure \ref{systemconfiguration}, where icons correspond to conversion or storage nodes, while bullets and arrows schematically represent conservation hyperedges. For the sake of readability, the set of nodes is split into three clusters, which also correspond to the different geographical areas displayed in Figure \ref{map_schematic}. The nodes and hyperedges used to model this system are described in the following subsections.

\subsubsection{Conversion Nodes}\label{conversionnodes}

Conversion nodes are discussed in this subsection. Tables \ref{conv_tech_params} and \ref{conv_econ_params} gather the techno-economic data (2030 estimates) used to model conversion nodes along with the original data sources and complement the descriptions below. In the model, power flows are measured in GW (GWh/h), energy is measured in GWh, mass flows are measured in kt/h, mass is measured in kt, and money is measured in M\euro.

\begin{table}[h!]
\caption{Technical parameters used to model conversion nodes.}
	\label{conv_tech_params}
	\resizebox{\textwidth}{!}{%
	\begin{tabular}{|c|ccccc|}
	\hline
	& $\phi_1$ & $\phi_2$ & $\phi_3$ & $\mu$ & $\Delta_{+,-}$ \\
	\hline
	HVDC Interconnection & \small{0.9499} & & & &\\
	\citep{Xiang2016,hvdclineefficiency} & \small{-} & & & &\\
	\hline
	Electrolysis & \small{50.6} & \small{9.0} & \small{8.0} & \small{0.05} & \small{1.0}\\
	\citep{Goetz2016} & \small{GWh$_{el}$/kt$_{H_2}$} & \small{kt$_{H_2O}$/kt$_{H_2}$} & \small{kt$_{O_2}$/kt$_{H_2}$} & \small{-} & \small{-/h} \\
	\hline
	Methanation & \small{0.5} & \small{2.75} & \small{2.25} & \small{1.0} & \small{0.0} \\
	\citep{Goetz2016,Roensch2016} & \small{kt$_{H_2}$/kt$_{CH_4}$} & \small{kt$_{CO_2}$/kt$_{CH_4}$} & \small{kt$_{H_2O}$/kt$_{CH_4}$} & \small{-} & \small{-/h}\\
	\hline
	Desalination & \small{0.004} & & & \small{1.0} & 0.0\\
	\citep{desalinationconsumption} & \small{GWh$_{el}$/kt$_{H_2O}$} & & & \small{-} & \small{-/h}\\
	\hline
	Direct Air Capture & \small{0.1091} & \small{0.0438} & \small{5.0} & \small{1.0} & \small{0.0} \\
	 \citep{Keith2018} &\small{GWh$_{el}$/kt$_{CO_2}$}&\small{kt$_{H_2}$/kt$_{CO_2}$} &\small{kt$_{H_2O}$/kt$_{CO_2}$}& \small{-} & \small{-/h} \\
	\hline
	CH$_4$ Liquefaction & \small{0.616} & & & \small{0.0} & 1.0\\
	\citep{Pospisil2019} & \small{GWh$_{el}$/kt$_{LCH_4}$} & & & \small{-}  & \small{-/h}\\
	\hline
	LCH$_4$ Carriers & \small{0.994} & & & &\\
	\citep{lngcarrierlifetime} & \small{-} & & & & \\
	\hline
	LCH$_4$ Regasification & \small{0.98} & & & &\\
	\citep{Pospisil2019} & \small{-} & & & &\\
	\hline
	\end{tabular}}
\end{table}

\begin{table}
\caption{Economic parameters used to model conversion nodes (2030 estimates).}
	\label{conv_econ_params}
	\resizebox{\textwidth}{!}{%
	\begin{tabular}{|c|cccc|}
	\hline
	& CAPEX & FOM (\small{$\theta_f$}) & VOM (\small{$\theta_v$}) & Lifetime\\
	\hline
	Solar Photovoltaic Panels & \small{380.0} & \small{7.25} & \small{0.0} & \small{25.0} \\
	\citep{solarpvcosts} & \small{M\euro/GW$_{el}$} & \small{M\euro/GW$_{el}$-yr} & \small{M\euro/GWh$_{el}$} & \small{yr}\\
	\hline
	Wind Turbines & \small{1040.0} & \small{12.6} & \small{0.00135} & \small{30.0}\\
	\citep{windcosts} & \small{M\euro/GW$_{el}$} & \small{M\euro/GW$_{el}$-yr} & \small{M\euro/GWh$_{el}$} & \small{yr}\\
	\hline
	HVDC Interconnection & \small{480.0} & \small{7.1} & \small{0.0} & \small{40.0}\\
	\citep{hvdccost,hvdccostsEIA} & \small{M\euro/GW$_{el}$} & \small{M\euro/GW$_{el}$-yr} & \small{M\euro/GWh$_{el}$} & \small{yr}\\
	\hline
	Electrolysis & \small{600.0} & \small{30.0} & \small{0.0} & \small{15.0} \\
	\citep{electrolysiscosts} & \small{M\euro/GW$_{el}$} & \small{M\euro/GW$_{el}$-yr} & \small{M\euro/GWh$_{el}$} & \small{yr}\\
	\hline
	Methanation & \small{735.0} & \small{29.4} & \small{0.0} & \small{20.0} \\
	\citep{methanationcosts} & \small{M\euro/GW$_{CH_4}$ (HHV)} & \small{M\euro/GW$_{CH_4}$-yr (HHV)} & \small{M\euro/GWh$_{CH_4}$ (HHV)} & \small{yr}\\
	\hline
	Desalination & \small{28.08} & \small{0.0} & \small{0.000315} & \small{20.0} \\
	\citep{desalinationcosts} & \small{M\euro/(kt$_{H_2O}$/h)} & \small{M\euro/(kt$_{H_2O}$/h)-yr} & \small{M\euro/kt$_{H_2O}$} & \small{yr}\\
	\hline
	Direct Air Capture & \small{4801.4} & \small{0.0} & \small{0.0207} & \small{30.0} \\
	\citep{Keith2018} &\small{M\euro/(kt$_{CO_2}$/h)} &\small{M\euro/(kt$_{CO_2}$/h)-yr} &\small{M\euro/kt$_{CO_2}$} & \small{yr}\\
	\hline
	CH$_4$ Liquefaction & \small{5913.0} & \small{147.825} & \small{0.0} & \small{30.0}\\
	\citep{liquefactioncosts} & \small{M\euro/(kt$_{LCH_4}$/h)} & \small{M\euro/(kt$_{LCH_4}$/h)-yr} & \small{M\euro/kt$_{LCH_4}$} & \small{yr}\\
	\hline
	LCH$_4$ Carriers & \small{2.537} & \small{0.12685} & \small{0.0} & \small{30.0} \\
	\citep{lngcarriercapex} & \small{M\euro/kt$_{LCH_4}$} & \small{M\euro/kt$_{LCH_4}$-yr} & \small{M\euro/kt$_{LCH_4}$} & \small{yr}\\
	\hline
	LCH$_4$ Regasification & \small{1248.3} & \small{24.97} & \small{0.0} & \small{30.0}\\
	\citep{regasificationcosts} & \small{M\euro/(kt$_{CH_4}$/h)} &\small{M\euro/(kt$_{CH_4}$/h)-yr} &\small{M\euro/kt$_{CH_4}$} & \small{yr}\\
	\hline
	\end{tabular}}
\end{table}

\begin{figure*}[h!]
\centering
\includegraphics[scale=0.75]{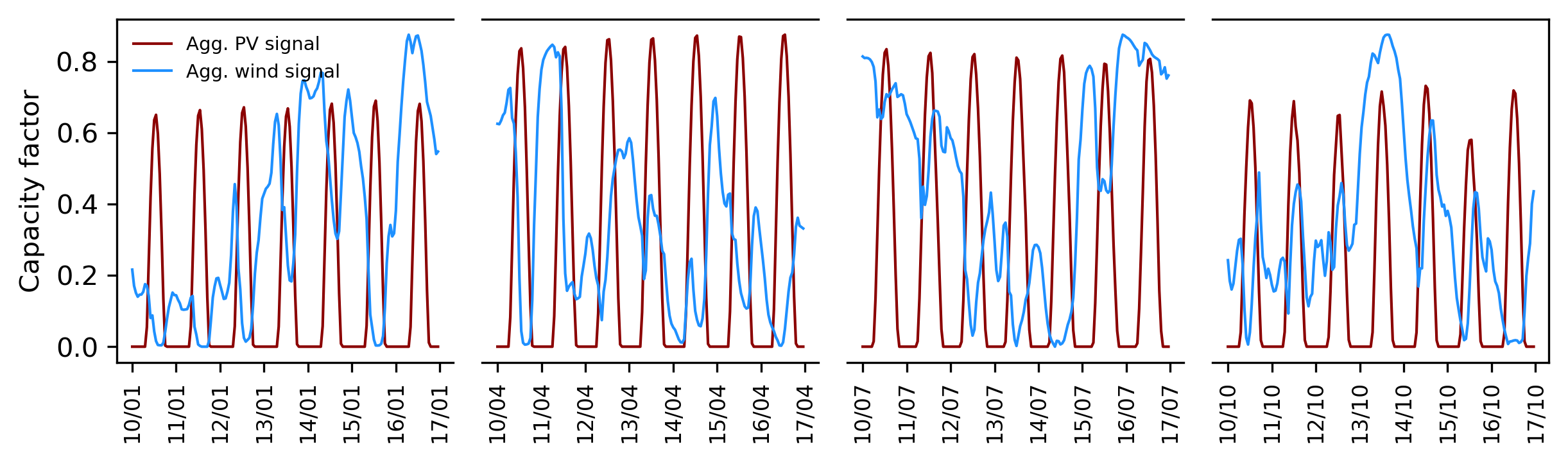}
\caption{Capacity factor time series $\pi_{t}^n$ used for solar PV and wind power plants in 2016. For all other nodes except liquefied methane carriers, $\pi_{t}^n = 1$ over the entire time horizon.}
\label{series_res}
\end{figure*}

\textbf{Solar PV} Solar photovoltaic panels are used for power generation. The plants are modelled with one external variable representing the output power and one internal variable representing the plant capacity, respectively. Constraints (\ref{sizing}) and (\ref{potential}) are used along with the local objective function \eqref{objectiveconversion}. In order to construct the capacity factor time series $\pi_{t}^n$, five years (2015-2019) of irradiance data at hourly resolution are retrieved from the ERA5 database \citep{ERA5} for each grey point in Figure \ref{map_schematic} and converted into capacity factors using a generic transfer function \citep{solarpvtransfer} and \textit{TrinaSolar Tallmax M} tilted module data \citep{trinasolar}. Sites with a five-year average capacity factor value exceeding $24.5$\% are retained (11 in total, shown by red crosses in Figure \ref{map_schematic}) and the associated time series are then aggregated (spatially averaged) into a single time series $\pi_{t}^n$, which is illustrated in Figure \ref{series_res} for a set of weekly periods in 2016.

\textbf{Wind Turbines} Wind turbines are used for generating power as well. Wind power plants are modelled in a similar fashion to solar PV plants, that is, with one external variable representing the power output and one internal variable representing the plant capacity, respectively. Constraints (\ref{sizing}) and (\ref{potential}) are used along with the typical local objective function \eqref{objectiveconversion}. In order to construct the capacity factor time series $\pi_{t}^n$, five years (2015-2019) of wind speed data at hourly resolution are retrieved from the ERA5 database \citep{ERA5} for each grey point in Figure \ref{map_schematic} and converted into capacity factors using the transfer function of the \textit{Vestas V90} turbine available in the \textit{windpowerlib} library \citep{windpowerlib}. Sites with a five-year average capacity factor value exceeding $50$\% are retained (5 in total, shown by blue crosses in Figure \ref{map_schematic}) and the associated time series are then aggregated (spatially averaged) into a single time series $\pi_{t}^n$, which is also displayed in Figure \ref{series_res}. 

\textbf{HVDC Interconnection} Ultra high voltage direct current (HVDC) overhead lines ($800$ or $1100$ kV) are assumed to be used for bulk power transmission from the first cluster (inland) to the second one (coastal hub) \citep{hvdccost}. Note that the yellow area containing the solar and wind sites in Figure \ref{map_schematic} is assumed to be a copper plate for the purpose of this study, which implies that solar PV and wind power plants feed directly into the electricity interconnection and the cost of the infrastructure connecting power plants to the HVDC interconnection is neglected. Voltage source converters (VSC) are well-suited for remote applications, as they are self-commutated and are much more controllable than typical line-commutated (LC) alternatives, although they are more expensive and have higher conversion losses \citep{Xiang2016}. In this case, two VSC stations are placed on each side of an overhead HVDC cable whose length is assumed to be $1000$ km. Losses in each converter station roughly amount to $1.8$\% of the power flowing through it, while approximately $1.5$\% of the power transiting through the HVDC cable is lost. Combining these figures yields the overall efficiency reported in Table \ref{conv_tech_params}. Economic data shown in Table \ref{conv_econ_params} include the costs of both converter stations and the cable. The interconnection is modelled using two external variables and one internal variable. The external variables represent the power flows into (i.e., leaving the first cluster) and out of (i.e., reaching the second cluster) the HVDC interconnection, respectively. The internal variable is the capacity of the converter-line pair. Investment costs in lines and converter stations are accounted for in the local objective \eqref{objectiveconversion}, along with operating costs.

\textbf{Electrolysis Plants} Proton exchange membrane (also called polymer electrolyte membrane, PEM) electrolysis plants \citep{Carmo2013} are used for producing hydrogen. This technology makes it possible to split water into hydrogen and oxygen by the passage of an electric current. Hence, the plants are modelled with four external and one internal variables. The external variables represent the power and water inflows as well as the hydrogen and oxygen outflows, while the internal variable is the plant capacity. The reference commodity $r$ used in Equation \eqref{conversion} is hydrogen, while the commodity $r'$ according to which the technology is sized in Equation \eqref{sizing} is the power input. Electrolysis plants are also assumed to operate at $20$ bar and $40$\degree C. This technology is flexible and can ramp up and down very quickly (usually within seconds). Hence, no ramping constraints are used. However, a minimum hydrogen production level around $5-10$\% of the nominal capacity must be maintained when the plant is switched on. Constraints \eqref{mustrun} are therefore used to model plant operation. The usual objective function \eqref{objectiveconversion} is also used. 

\textbf{Methanation Plants} The carbon dioxide methanation (Sabatier) reaction enables the conversion of carbon dioxide and hydrogen into methane and water (steam) and is highly exothermic (i.e., the production of $1$ kg of methane releases approximately $2.867$ kWh of high-temperature heat \citep{Roensch2016}). In this paper, cooled fixed-bed (catalytic) reactors operating at $300$\degree C and $20$ bar are assumed to be used to produce synthetic methane via the Sabatier reaction. Furthermore, carbon dioxide and hydrogen are assumed to be fed in stoichiometric proportions, and the conversion of reactants is assumed to be complete (which is facilitated by the use of, e.g., alumina-supported nickel catalysts \citep{Mills1974} that offer good selectivity and are relatively cheap). Plants are modelled using four external variables and one internal variable. The external variables represent the hydrogen and carbon dioxide inflows as well as the methane and water (steam) outflows, while the internal variable is the plant capacity. Methane is taken as the reference commodity $r$ used to describe the process in Equation \eqref{conversion} as well as the reference commodity $r'$ used for sizing the plant in Equation \eqref{sizing}. Owing to the exothermicity of the reaction, cooled fixed-bed reactors are very sensitive to changes in operating parameters such as the feed and coolant temperatures \citep{Schlereth2014} or the feed flow rate \citep{Theurich2019}, and can suffer from pronounced hot-spots or thermal runaway \citep{Schlereth2014}, which can in turn lead to catalyst sintering and deactivation. Hence, reactors usually have a (very) limited operating range, although promising ways of improving this have been proposed \citep{Bremer2019}. Dynamic operation may also involve shutting down the reactor, keeping it in hot standby and starting it up again \citep{Gorre2020}, which nevertheless leads to inefficiencies (e.g., the reactor must be flushed with hydrogen \citep{Gorre2020}). Finally, note that maintaining product quality is typically more difficult in unsteady operation. In the light of these observations, in this paper, it is assumed that methanation reactors operate in steady state. Constraints \eqref{mustrun}, \eqref{rampup}, \eqref{rampdown} are therefore used to model plant operation, while investment and operating costs are modelled via \eqref{objectiveconversion}.

\textbf{Water Desalination Plants} Reverse osmosis (RO) plants are employed to desalinate seawater and produce freshwater \citep{Caldera2016}. This technology essentially pumps seawater into a chamber featuring a porous membrane and produces a pressure differential across the membrane, enabling dead-end filtration and the recovery of freshwater on the other side of the membrane. The plants are modelled with two external variables and one internal variable. The external variables are the power required to drive pumps and the freshwater outflow, whereas the internal variable represents the plant capacity. The reference commodity $r'$ according to which the plant is sized is the freshwater flow out of the system. For mechanical reasons, membranes are usually designed to operate under constant pressure and plants therefore operate more or less continuously. Hence, constraints \eqref{conversion},\eqref{sizing},\eqref{mustrun}, \eqref{rampup}, \eqref{rampdown} are used to model plant sizing and operation, while investment and operating costs are modelled via \eqref{objectiveconversion}. Note that the seawater inflow and brine discharge are not modelled. The implicit assumptions are that seawater is freely available and that the brine by-product can be disposed of at no cost, without any restriction on pumped volumes.

\textbf{Direct Air Capture Units} Direct air capture units extract carbon dioxide from the atmosphere \citep{Kiani2020}. The process used in this paper is the one proposed by \cite{Keith2018}. Roughly speaking, this process relies on four main chemical reactions, which are combined to form two chemical loops. In the first loop, aqueous sorbents are used in an air contactor to chemically bind carbon dioxide and form dissolved compounds. These compounds then react with pellets in a fluidised-bed reactor, making it possible to recover the aforementioned sorbents and trap carbon in solid compounds. The second loop essentially recovers carbon dioxide by calcining the solid compounds and replenishes the pellet stock by hydrating (slaking) the solid product of the calcination reaction. The process requires electricity to power fans driving air through the contactors, pumps maintaining the flow of aqueous solutions as well as compressors compressing the output carbon dioxide stream from atmospheric pressure to $20$ bar (the associated energy expense is approximated via the polytropic compression work, assuming a polytropic efficiency of $80$\%). The net power consumption is obtained as the difference between the total consumption of these subsystems and the power produced by a steam turbine recovering slaking heat. A sustained water supply is also necessary to form aqueous solutions, counter natural evaporation in the air contactors and produce steam used in the slaker. Furthermore, a source of heat at around 900\degree C is required for the calcination reaction. In the original design, natural gas is burnt via an oxy-fuel combustion process at the bottom of the calciner to provide this heat, and the off-gases fluidise the reactor. In this paper, it is assumed that the high temperature heat (approximately $1.46$ MWh per ton of carbon dioxide) is provided by burning hydrogen (assuming a lower heating value of $33.3$ MWh per ton of hydrogen burnt). Hence, the process is modelled using four external variables and one internal variable. The external variables represent the power, water and hydrogen inflows as well as the carbon dioxide outflow, while the internal variable is the plant capacity. The reference flow according to which the plant is sized is the carbon dioxide outflow. None of the technologies implementing the various reactions really lend themselves to highly variable operation. Constraints \eqref{conversion},\eqref{sizing},\eqref{mustrun}, \eqref{rampup}, \eqref{rampdown} are therefore used to model plant sizing and operation, while investment and operating costs are modelled via \eqref{objectiveconversion}.

\textbf{Methane Liquefaction Units} Liquefaction units turn gaseous methane into liquefied methane \citep{Pospisil2019}. This technology typically relies on compressors and pumps in order to progressively compress and cool the methane inflow, which is eventually throttled and liquefied via the Joule-Thomson effect. In this case, three external variables and one internal variable are used. The external variables represent the methane inflow, the power consumption of compressors and pumps as well as the liquefied methane outflow (which is the reference commodity), while the internal variable represents the plant capacity. Constraints \eqref{conversion},\eqref{sizing},\eqref{mustrun}, \eqref{rampup}, \eqref{rampdown} are used to model plant sizing and operation, while investment and operating costs are modelled via \eqref{objectiveconversion}.

\begin{figure*}[h!]
\centering
\includegraphics[scale=0.75]{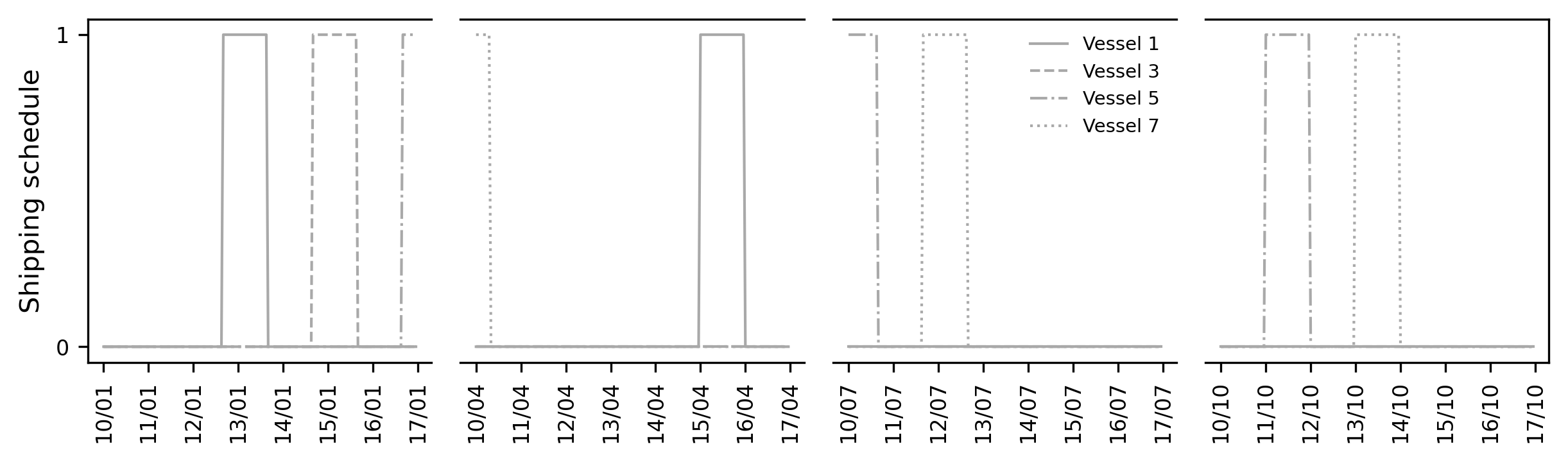}
\caption{Subset of non-overlapping schedules used to construct the aggregate schedule $\pi_t^n$ of stylised liquefied methane carriers. These time series are summed to obtain the aggregate schedule $\pi_t^n$. For all other nodes except wind and solar PV power plants, $\pi_{t}^n = 1$ over the entire time horizon.}
\label{series_ship}
\end{figure*}

\textbf{Liquefied Methane Carrier Vessels} Liquefied methane is transported to market with large ocean-going vessels powered by dual fuel diesel electric (DFDE) engines \citep{lngcarrierlifetime}. These engines are particularly efficient and can run solely on natural boil-off gas (i.e., gaseous methane resulting from the natural evaporation of liquefied methane stored on board in insulated cargo tanks). This allows vessels to sail at a speed of 19 knots, with approximately 0.1\% of their cargo evaporating due to natural boil-off per day spent at sea, which is used for propulsion (i.e., no other fuel is needed). The liquefied methane heel that must usually be maintained for the return journey to guarantee that the onboard tanks remain cool (roughly 4-5\% of the total cargo) is neglected in this paper. Two external variables and one internal variable are used to describe a stylised carrier vessel. The external variables represent the flow of liquefied methane loaded at the coastal hub and the flow of liquefied methane unloaded at the destination, respectively. The internal variable is the vessel capacity. Eq. \eqref{conversion} is used to model the transport of liquefied methane, with $\tau = 116$ hours, as the berthing and travel time between the coastal hub and the destination is assumed to take slightly less than 5 days. The conversion factor $\phi \approx 0.994 $ represents the transport efficiency, computed from the boil-off consumption (0.125\% of cargo per day) and trip duration (116 hours). In addition, loading and unloading may only be possible when the vessel is moored at the coastal hub and destination, respectively. This is enforced via Eq. \eqref{sizing} and time series $\pi_{t}^n$ (with values equal to 0 or 1), which defines a berthing, mooring, loading and unloading schedule (loading or unloading take place when $\pi_{t}^n = 1$). For the sake of simplicity, $\pi_{t}^n$ represents an aggregate schedule constructed from 7 different, non-overlapping schedules corresponding to individual carrier vessels. Some of these schedules are shown in Figure \ref{series_ship} (loading and unloading is assumed to take 24 hours). The standard local objective \eqref{objectiveconversion} is used for the stylised carrier.

\textbf{Liquefied Methane Regasification Units} Regasification units are used to transform liquefied methane into gaseous methane at the destination \citep{regasificationcosts}. The heat required to do so can come from a variety of sources. In this case, it is assumed to come from the combustion of 	a fraction of the methane (around 2\%). Thus, two external variables and one internal variable are used. The external variables represent the liquefied methane inflow as well as the gaseous methane outflow, and the internal variable is the plant capacity. Constraints \eqref{conversion},\eqref{sizing} are used to model plant sizing and operation, while investment and operating costs are modelled via \eqref{objectiveconversion}.

\subsubsection{Storage Nodes}\label{storagenodes}

Storage nodes are discussed in this subsection. Tables \ref{stor_tech_params}, \ref{stor_econ_stock_params} and \ref{stor_econ_flow_params} gather the techno-economic data (2030 estimates) used to model storage nodes along with the original data sources and complement the descriptions below. In the model, power flows are measured in GW (GWh/h), energy is measured in GWh, mass flows are measured in kt/h, mass is measured in kt, and money is measured in M\euro.

\begin{table}
	\caption{Technical parameters used to model storage nodes.}
	\label{stor_tech_params}
	\resizebox{\textwidth}{!}{%
	\begin{tabular}{|c|cccccc|}
		\hline
		&  \small{$\eta^S$} & \small{$\eta^+$} & \small{$\eta^-$} & \small{$\sigma$} & \small{$\rho$} & \small{$\phi$} \\ \hline
		Battery Storage & \small{0.00004} & \small{0.959} & \small{0.959} & \small{0.0} & \small{1.0} & \\
		\citep{batterycosts} & \small{-} & \small{-} & \small{-} & \small{-} & \small{-} &\\
		\hline
		Compressed H$_2$ Storage & \small{1.0} & \small{1.0} & \small{1.0} & \small{0.05} & \small{1.0} & \small{1.3}\\
		\citep{batterycosts} & & & & & & \small{GWh$_{el}$/kt$_{H_2}$}\\
		\hline
		Liquefied CO$_2$ Storage & \small{1.0} & \small{1.0} & \small{1.0} & \small{0.0} & \small{1.0} & \small{0.105}\\
		\citep{co2liquefaction} & & & & & & \small{GWh$_{el}$/kt$_{CO_2}$}\\
		\hline
		Liquefied CH$_4$ Storage & \small{1.0} & \small{1.0} & \small{1.0} & \small{0.0} & \small{1.0} & \\
		 & \small{-} & \small{-} & \small{-} &\small{-} & \small{-} & \\
		\hline
		H$_2$O Storage & \small{1.0} & \small{1.0} & \small{1.0} & \small{0.0} & \small{1.0} & \small{0.00036}\\
		\citep{Caldera2016} & & & & & &\small{GWh$_{el}$/kt$_{H_2O}$}\\
		\hline
	\end{tabular}}
\end{table}

\begin{table}
	\caption{Economic parameters used to model storage nodes (stock component, 2030 estimates).}
	\label{stor_econ_stock_params}
	\centering
	\resizebox{\textwidth}{!}{%
	\begin{tabular}{|c|cccc|}
		\hline
		& CAPEX & FOM (\small{$\vartheta_f$})& VOM (\small{$\vartheta_v$}) & Lifetime \\ 
		\hline
		Battery Storage & \small{142.0} & \small{0.0} & \small{0.0018} & \small{10.0}\\
		 \citep{batterycosts} & \small{M\euro/GWh} & \small{M\euro/GWh-yr} & \small{M\euro/GWh} & \small{yr} \\
		\hline
		Compressed H$_2$ Storage & \small{45.0} & \small{2.25} & \small{0.0} & \small{30.0}\\
		 \citep{batterycosts} & \small{M\euro/kt} & \small{M\euro/kt-yr} & \small{M\euro/kt} & \small{yr}\\
		\hline
		Liquefied CO$_2$ Storage & \small{1.35} & \small{0.0675} & \small{0.0} & \small{30.0}\\
		\citep{co2liquefaction} & \small{M\euro/kt} & \small{M\euro/kt-yr} & \small{M\euro/kt} & \small{yr} \\
		\hline
		Liquefied CH$_4$ Storage & \small{2.641} & \small{0.05282} & \small{0.0} & \small{30.0}\\
		 \citep{lngtankcosts} & \small{M\euro/kt}& \small{M\euro/kt-yr}& \small{M\euro/kt} & \small{yr} \\
		\hline
		H$_2$O Storage & \small{0.065} & \small{0.0013} & \small{0.0} & \small{30.0}\\
		 \citep{Caldera2016} & \small{M\euro/kt} & \small{M\euro/kt-yr} & \small{M\euro/kt} & \small{yr}\\
		\hline
	\end{tabular}}
	\end{table}
\begin{table}
	\centering
	\caption{Economic parameters used to model storage nodes (flow component, 2030 estimates).}
	\label{stor_econ_flow_params}
	\resizebox{\textwidth}{!}{%
	\begin{tabular}{|c|cccc|}
		\hline
		& CAPEX & FOM (\small{$\theta_f$}) & VOM (\small{$\theta_v$}) & Lifetime \\ 
		\hline
		Battery Storage & \small{160.0} & \small{0.5} & \small{0.0} & \small{10.0} \\
		 \citep{batterycosts} & \small{M\euro/GW} & \small{M\euro/GW-yr} & \small{M\euro/GWh} & \small{yr}\\
		\hline
		Liquefied CO$_2$ Storage & \small{48.6} & \small{2.43} & \small{0.0} & \small{30.0}\\
		\citep{co2liquefaction} & \small{M\euro/(kt/h)} & \small{M\euro/(kt/h)} & \small{M\euro/kt} & \small{yr}\\
		\hline
		H$_2$O Storage & \small{1.55923} & \small{0.0312} & \small{0.0} & \small{30.0}\\
		 \citep{Caldera2016} & \small{M\euro/(kt/h)} & \small{M\euro/(kt/h)} & \small{M\euro/kt} & \small{yr}\\
		\hline
	\end{tabular}}
\end{table}

\textbf{Stationary Battery Storage} Nickel manganese cobalt (NMC) oxide lithium-ion batteries are used for short-term electricity storage \citep{batterycosts}. Power in- and outflows are modelled using external variables. The state of charge, power capacity and energy capacity, on the other hand, are modelled as internal variables. Constraints \eqref{storagedynamics}, \eqref{storagecyclicity}, \eqref{storagestocksizing}, \eqref{storagepotential}, \eqref{storageflowplussizing}, \eqref{storageflowminussizing} are used, while the local objective function is given in \eqref{objectivestorage}.

\textbf{Hydrogen Storage Tanks} Compressed hydrogen storage tanks are considered in this paper. More precisely, 
overground, man-made steel storage vessels (type I) withstanding pressure levels around 200 bar and suitable for stationary applications are used \citep{batterycosts}. A minimum inventory level of 5\% is assumed to represent the cushion gas (effectively reducing the working volume). Since hydrogen at 20 bar and 40\degree C is produced by electrolysis plants, the hydrogen inflow must be compressed to 200 bar using electric compressors for storage purposes. The associated energy expense is approximated via the polytropic compression work (assuming a polytropic efficiency of 80\%) \citep{JRC2003}. Thus, three external variables and three internal variables are used. The external variables represent the hydrogen inflow, the electricity consumption and the hydrogen outflow, while the state of charge, the power capacity and the energy capacity are modelled as internal variables. Constraints \eqref{storagedynamics}, \eqref{conversionstorage}, \eqref{storagecyclicity}, \eqref{storagestocksizing},  \eqref{storagepotential}, \eqref{storageflowplussizing}, \eqref{storageflowminussizing} are used, while the local objective function is \eqref{objectivestorage}.

\textbf{Liquefied Carbon Dioxide Storage Tanks} Liquefied carbon storage tanks are used to store carbon dioxide. Liquefaction and regasification units are also required \citep{co2liquefaction}. Liquefaction units consume electricity, while regasification units are assumed to use ambient heat to recover gaseous carbon dioxide. Hence, in this case, three external variables and five internal variables are used. The external variables are the gaseous carbon dioxide inflow, the power consumption of the liquefaction units and the gaseous carbon dioxide outflow, while the internal variables represent the state of charge, the tank capacity, capacities of liquefaction and regasification units, and the flows of liquefied carbon dioxide in and out of the tanks. Constraints \eqref{storagedynamics}, \eqref{conversionstorage}, \eqref{storagecyclicity}, \eqref{storagestocksizing}, \eqref{storagepotential}, \eqref{storageflowplussizing}, \eqref{storageflowminussizing} are used, while the local objective function is \eqref{objectivestorage}.

\textbf{Liquefied Methane Storage Tanks} Liquefied methane is stored in full containment tanks (i.e., tanks with both inner and outer containment walls and such that the annular gap between both walls is sealed to prevent any gaseous leaks \citep{lngtankcosts}). It is assumed that the boil-off gas keeping the content of the storage tanks cold is re-liquefied and pumped back into the tanks but the electricity consumption required to do so is neglected. Two external variables and two internal variables are used. The external variables are the liquefied methane in- and outflow, while internal variables represent the state of charge and the storage capacity. Constraints \eqref{storagedynamics}, \eqref{storagecyclicity}, \eqref{storagestocksizing}, \eqref{storagepotential}, \eqref{storageflowplussizing}, \eqref{storageflowminussizing} are used, while the local objective function is \eqref{objectivestorage}.

\textbf{Water Storage Tanks} Water is stored in tanks equipped with electric pumps \citep{Caldera2016}. Three external variables and three internal variables are used. The external variables correspond to the water inflow, the power consumed by pumps and the the water outflow. The internal variables represent the state of charge, the tank capacity and the flow capacity of pipes feeding into the tank. Constraints \eqref{storagedynamics}, \eqref{storagecyclicity}, \eqref{storagestocksizing}, \eqref{storagepotential}, \eqref{storageflowplussizing}, \eqref{storageflowminussizing} are used, while the local objective function is \eqref{objectivestorage}.

\subsubsection{Conservation Hyperedges}
 
\textbf{Inland Power Balance} This hyperedge enforces active power flow conservation (which derives from Kirchhoff's current law) in the inland cluster. It therefore guarantees that the sum of power flows from the solar PV plant, the wind power plant and the battery is equal to the sum of power flows to the HVDC interconnection and the battery. Note that both in and outflows are used for the battery, which correspond to discharge and charge flows, respectively. No exogenous power injections and withdrawals take place over this hyperedge, hence $\lambda_t^e = 0, \forall t \in \mathcal{T}$. This is assumed to be the case for all other hyperedges as well, unless otherwise stated.

\textbf{Coastal Power Balance} This hyperedge enforces active power flow conservation in the coastal cluster. It therefore guarantees that the power flow from the HVDC interconnection is equal to the sum of power flows to the direct air capture plant, the electrolysis plant, the hydrogen storage system, the methane liquefaction units, the desalination plant and the liquefied carbon dioxide storage system.

\textbf{Coastal Hydrogen Balance} This hyperedge enforces conservation of hydrogen mass flows in the coastal cluster. It therefore guarantees that the sum of flows from the electrolysis plants and the storage system is equal to the sum of flows to the direct air capture plants, the methanation plants and the storage system.

\textbf{Coastal Carbon Dioxide Balance} This hyperedge enforces conservation of (gaseous) carbon dioxide mass flows in the coastal cluster. It therefore guarantees that the sum of flows from the direct air capture units and the storage system is equal to the sum of flows to methanation plants and the storage system.

\textbf{Coastal Water Balance} This hyperedge guarantees that the aggregate flow of freshwater generated by desalination and methanation plants in the coastal cluster exceeds the aggregate flow consumed by electrolysis and direct air capture plants. Hence, it is assumed that any freshwater surplus may be released into the environment without harm or used in other applications (e.g., cooling), and the equality constraint in Eq. \eqref{conservationhyperedge} is relaxed to a $\ge$ inequality.

\textbf{Coastal Methane Balance} This hyperedge enforces conservation of (gaseous) methane mass flows in the coastal cluster, and guarantees that flows from the methanation plants and to the liquefaction units are equal.

\textbf{Coastal Liquefied Methane Balance} This hyperedge enforces conservation of liquefied methane mass flows in the coastal cluster. Thus, it guarantees that the sum of flows from the liquefaction units and the storage system is equal to the sum of flows to the storage system and the liquefied methane carriers.

\textbf{Destination Liquefied Methane Balance} This hyperedge enforces conservation of liquefied methane mass flows at the destination. Hence, it guarantees that the sum of flows from the liquefied methane carriers and the storage system is equal to the sum of flows to the regasification units and storage system.

\textbf{Destination Methane Balance} This hyperedge guarantees that the exogenous demand for methane at the destination is satisfied by the flows from the regasification units. The gas demand is set to $10$ TWh (HHV) per annum, but no specific assumptions about the end-uses or sectors relying on synthetic methane are made. Furthermore, the liquefied methane terminal located in Northwestern Europe is assumed to be connected to some existing gas network infrastructure. Although the latter is not explicitly modelled, it is assumed to be able to absorb both short-term and seasonal demand variability (e.g., via its line pack and seasonal storage facilities \citep{CorreaPosada2015}, as is the case in most systems). The demand profile, which essentially represents injections from the terminal into the gas network, is therefore assumed to be flat. Hence, assuming that synthetic methane has a HHV of $15.441$ kWh/kg, the demand profile is obtained as $\lambda_{t}^e = (10 \times 10^3/8760) \times (1/15.441) \approx 0.07393$ kt/h, $\mbox{ } \forall t \in \mathcal{T}$. Note that using the LHV would have resulted in a higher mass flow rate.
 
\subsection{Scenarios}
One reference scenario and a comprehensive sensitivity analysis are presented. The reference scenario studies a system configuration that relies on a combination of solar PV and wind power plants for electricity generation. A uniform weighted average cost of capital (WACC) of 7\% is assumed for all technologies, which represents the case where the funds required to finance the system are borrowed on capital markets. Under these assumptions, for a conversion or storage technology $n$, the CAPEX values in Tables \ref{conv_econ_params}, \ref{stor_econ_stock_params} and \ref{stor_econ_flow_params} are used to compute
\begin{equation}
\zeta^n = \mbox{CAPEX}_n \times \frac{\mbox{w}}{(1 - (1 + \mbox{w})^{-\mbox{L}_n})},
\label{waccformula}
\end{equation}
with $\mbox{L}_n$ the lifetime of technology $n$ and $\mbox{w}$ the WACC. Hence, $\zeta^n$ represents the annualised cost of investing in technology $n$.

The sensitivity analysis investigates the impact of a number of techno-economic parameters and assumptions on synthetic methane cost, which include the availability of wind power plants, the investment costs of electrolysis, direct air capture and methanation plants, the operational flexibility of the latter two technologies as well as the energy consumption of direct air capture plants and the financing costs. More specifically, a hypothetical situation where the cost of financing the system is zero is studied, such that the cost of synthetic methane production and delivery solely reflects the cost and efficiency of technologies in the supply chain. In this set-up, Eq. \eqref{waccformula} cannot be used and annualised investment costs are instead computed via
\begin{equation}
\zeta^n = \frac{\mbox{CAPEX}_n}{\mbox{L}_n}.
\label{nowaccformula}
\end{equation}

\section{Results}\label{results}
\subsection{Reference Scenario}\label{results_ref_scen}
In the reference scenario, a system configuration relying on solar and wind power plants is studied assuming a WACC of $7$\%. In this set-up, synthetic methane is delivered to market in gaseous form at $149.7$ \euro/MWh (HHV), which is computed as the ratio of total (annualised) system cost to methane volume delivered ($10$ TWh HHV per year). It is worth noting that using the LHV would have increased the cost per MWh, as this would have effectively reduced the amount of energy that could have been retrieved per unit mass of methane delivered.

\begin{figure}[h!]
\centering
\includegraphics[scale=0.45]{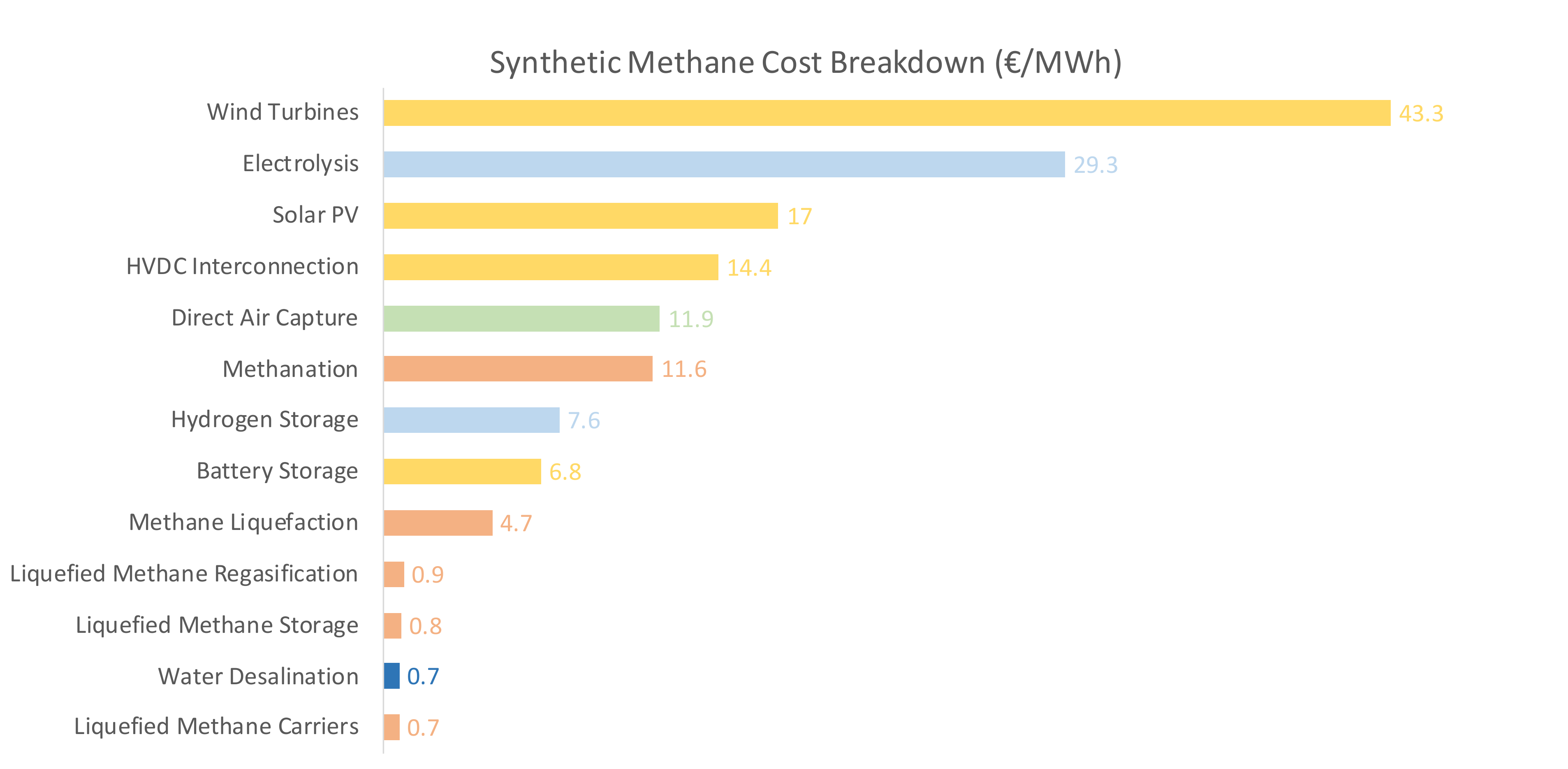}
\caption{Breakdown of synthetic methane cost at destination for reference scenario. All contributions roughly sum to $149.7$ \euro/MWh (HHV).}
\label{refscencosts}
\end{figure}

The synthetic methane cost breakdown is provided in Figure \ref{refscencosts}, where each bar represents the contribution (in \euro/MWh) of the corresponding technology to synthetic methane cost. Each bar can also be interpreted as representing the contribution of the corresponding technology to total system cost. Wind turbines account for the lion's share of synthetic methane cost (roughly $28.9\%$). Electrolysis ($19.6\%$) and solar PV plants ($11.3\%$) come in second and third, respectively, although they each represent a much smaller proportion of costs (taken together, they contribute slightly more than wind turbines to total system cost). Overall, the technologies used to generate, transport and store electricity (shown in gold in Figure \ref{refscencosts}) represent the largest share of costs (around $56.6\%$). Hydrogen storage plants, which are used as a buffer between flexible electrolysis and inflexible methanation plants, make up approximately $5\%$ of total cost. Hence, the technologies producing and storing hydrogen (shown in light blue in Figure \ref{refscencosts}) account for roughly $25\%$ of total system cost. It is worth noting that the plants upstream of the inflexible plants (i.e., methanation, direct air capture and desalination plants) make up almost $80\%$ of total system cost. On the other hand, methanation plants make up a minor share of total cost (approximately $7.7\%$), and the full methane chain (i.e., production, liquefaction, storage, transport and regasification, shown in light orange in Figure \ref{refscencosts}) accounts for roughly $12.5\%$ of final product cost. Direct air capture plants also represent a minor fraction of system cost (around $7.9\%$, shown in green in Figure \ref{refscencosts}). Water desalination and storage technologies are deployed in moderate quantities, resulting in a very small share of total costs (well under 1\%), while carbon dioxide storage is not deployed.

\begin{figure}[h!]
\centering
\includegraphics[scale=0.75]{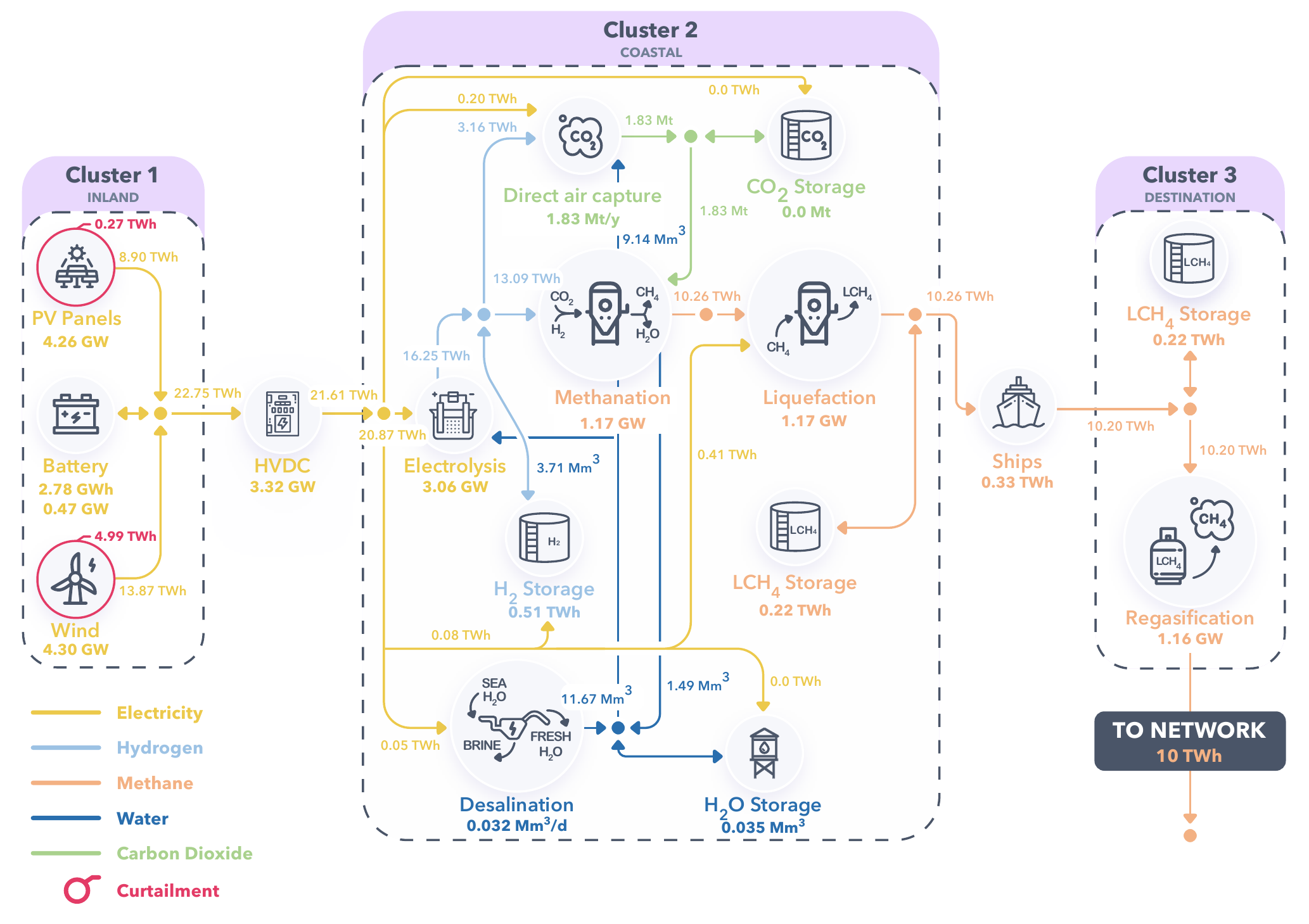}
\caption{Material and energy balance diagram for the reference scenario, along with technology capacities. All energy-equivalent flows of energy carriers other than electricity have been computed using their HHV. Flow values represent yearly averages (i.e., flows have been integrated over the full time horizon and divided by the number of years) and all values have been rounded up to keep significant digits only.}
\label{refscencapas}
\end{figure}

Analysing mass and energy balances provides some insight into system design and operation. Figure \ref{refscencapas} displays mass and energy balances (flow values are integrated over the full optimisation horizon of one year) along with technology capacities. Firstly, as can be seen in Figure \ref{refscencapas}, the average annual electricity production of solar PV and wind power plants is equal to $8.90$ and $13.87$ TWh, respectively, which suggests that the full supply chain has a conversion efficiency of roughly $43.9$\%. However, the amount of curtailment is substantial and stands at $5.26$ TWh, which represents slightly less than one quarter of the useful power production. This effectively decreases the capacity factors of the photovoltaic and wind power plants from their theoretical maximum of $24.6$\% and $50.0$\% over 2015-2019 (i.e., corresponding to the case where all electricity produced is used) to $23.8$\% and $36.8\%$ (taking only useful power production into account). However, the high share of curtailment attributed to wind power plants must be taken with caution. Indeed, since curtailment is not penalised in the objective, some natural symmetry exists in the model when solar PV and wind power plants jointly produce more than the system can absorb (which typically occurs during hours of peak solar PV production). More specifically, in such situations, solutions curtailing a given share of solar PV output would yield the same objective values as that of other solutions curtailing the same given share of wind power output, and curtailment could therefore be attributed to one technology or the other without any impact on the objective. The high overall rate of curtailment can nevertheless be explained by the difficulty of effectively absorbing the highly variable aggregate power input from renewable power plants. This is a direct consequence of the fact that the operating regimes of several key conversion technologies are inflexible, which has two further implications. Firstly, battery and hydrogen storage systems are deployed at great cost in order to smooth the variability of the power supply as much as possible. Secondly, plants located upstream of the inflexible ones are typically oversized (especially solar PV power plants), as the level of smoothing required to guarantee steady power and hydrogen flows cannot be economically provided by storage plants alone. Additional evidence supporting this analysis is provided in the following section.

\subsection{Sensitivity Analyses}\label{results_sensitivity}

\begin{figure}[h!]
\centering
\includegraphics[scale=0.5]{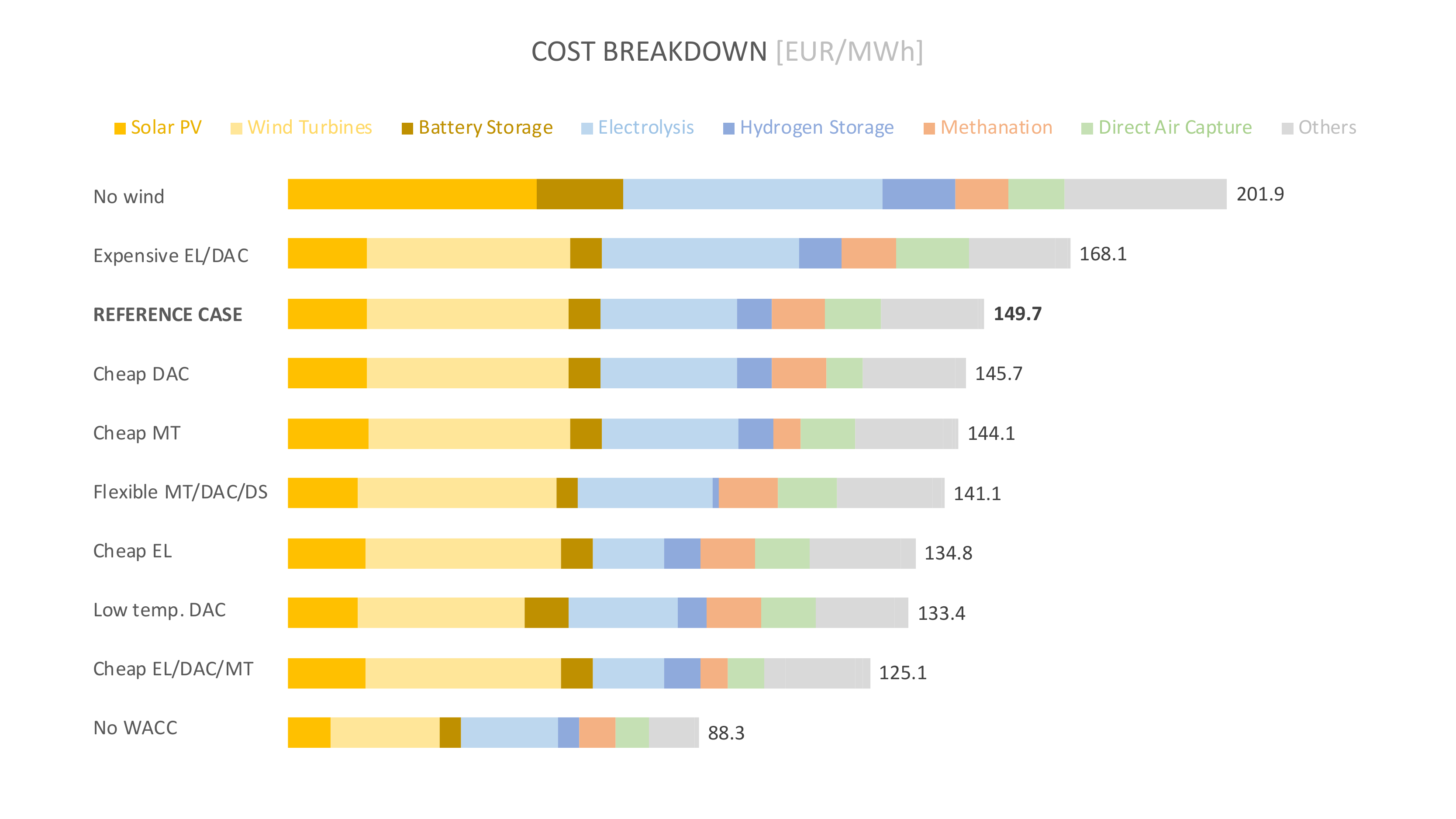}
\caption{Breakdown of synthetic methane costs obtained under various techno-economic assumptions.}
\label{sensitivitycosts}
\end{figure}

\begin{figure}[h!]
\centering
\includegraphics[scale=0.5]{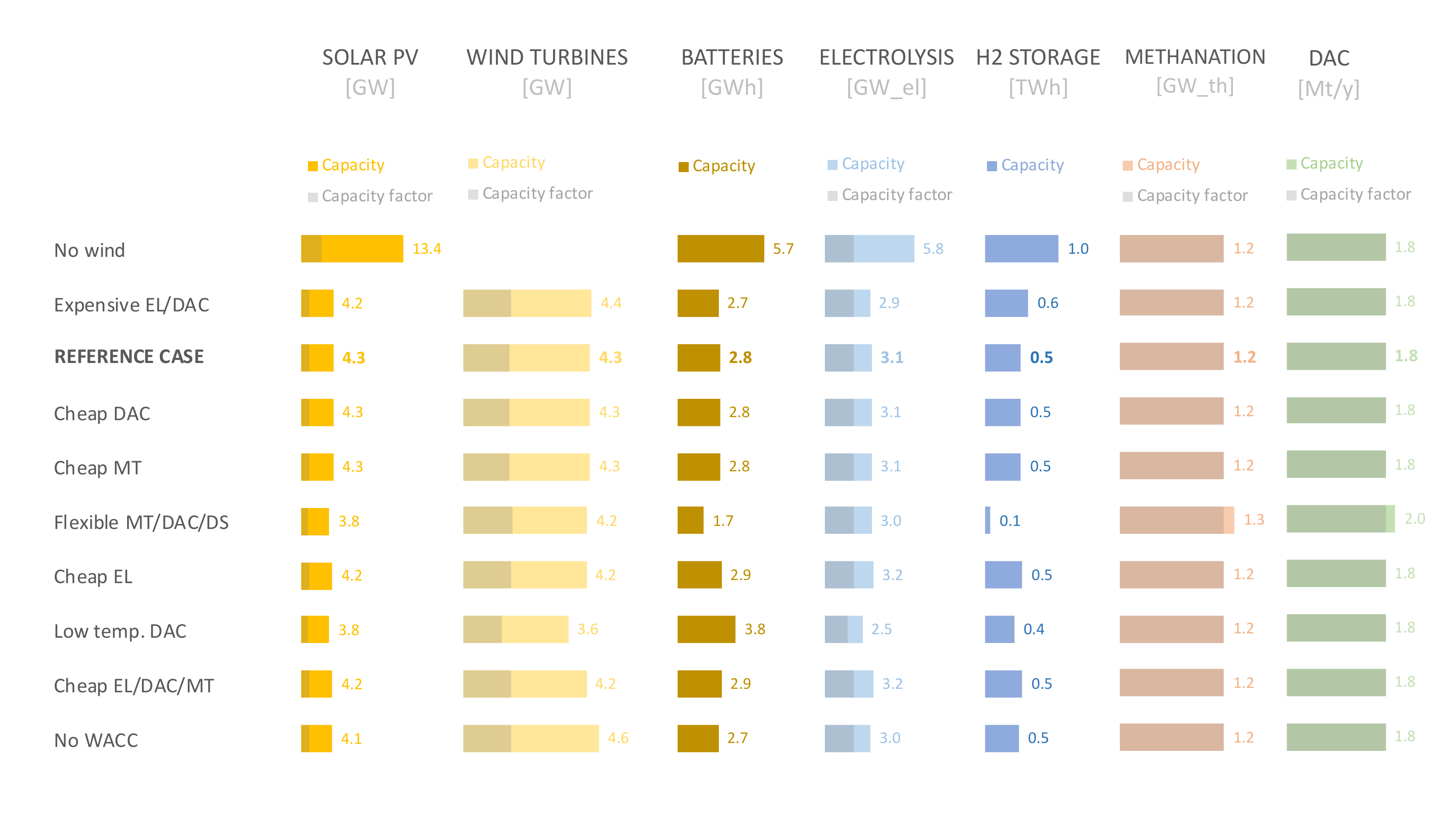}
\caption{Capacities and average capacity factors (shown as greyed fractions of capacity bars) of key conversion and storage technologies under various techno-economic assumptions.}
\label{sensitivitycapacities}
\end{figure}

The sensitivity analysis investigates the impact of a number of techno-economic parameters and assumptions on synthetic methane cost. More precisely, the impacts of i) being unable to deploy wind power plants ii) the operational flexibility of direct air capture, methanation and desalination plants iii) the investment costs of electrolysis, direct air capture and methanation plants iv) the energy consumption of direct air capture plants v) financing costs are assessed. The cost share, capacities and capacity factors of the technologies that were found to contribute the most to total system cost in Section \ref{results_ref_scen} are computed and analysed in this section. Hence, Figure \ref{sensitivitycosts} displays the breakdowns of synthetic methane costs obtained under various techno-economic assumptions, while Figure \ref{sensitivitycapacities} gathers the capacities and average capacity factors of key conversion and storage technologies. These results are elaborated upon below.

\textbf{Solar PV System} This analysis assumes that wind power plants cannot be deployed, and the full electricity supply must therefore come from solar PV power plants. As can be seen in Figure \ref{sensitivitycosts}, the cost of synthetic methane for this configuration is around $202$ \euro/MWh, which is almost $35\%$ more expensive than that found in the reference scenario. In this case, electrolysis plants contribute the most to total system cost, followed by solar PV power plants. In addition, it is clear from Figure \ref{sensitivitycosts} that the cost of each technology located upstream of the inflexible plants (i.e., methanation, direct air capture and desalination plants) increases in absolute terms compared with the reference scenario. The only technologies whose cost remains the same are direct air capture and methanation. This claim is supported by the results shown in Figure \ref{sensitivitycapacities}. Indeed, the capacities of conversion and storage technologies located upstream of the inflexible ones are larger in the solar-only case, while the capacities of inflexible technologies are equal in both cases. Furthermore, the average capacity factors of key conversion technologies (e.g., electrolysis plants) are much lower in the solar-only case. This observation suggests that these plants are oversized to help absorb the highly variable input from solar PV power plants, as smoothing it with storage systems alone would be uneconomical. Overall, relying on solar PV power plants alone results in a system design that it is much less efficient and much more expensive than the one identified in the reference scenario.

\textbf{System Flexibility} In the reference scenario, several technologies were assumed to be inflexible, namely methanation (MT), direct air capture (DAC) and desalination (DS) plants. Their inflexibility, combined with the fact that the renewable power supply is highly variable, was found to have an impact on synthetic methane cost. Although these assumptions are well-founded, the minimum level and ramping constraints of the three aforementioned technologies are now relaxed (i.e., $\mu = 0.0$ and $\Delta_+ = \Delta_- = 1.0$) in order to evaluate the sensitivity of our results. Figure \ref{sensitivitycosts} shows that shifting to a system with fully flexible methanation, direct air capture and desalination plants can lead to cost savings around $6$\%, and has a substantial impact on the capacities of several technologies. More specifically, the capacity of solar PV power plants decreases by $10$\%, while the capacities of battery and hydrogen storage systems shrink by $40$\% and $80$\%, respectively. On the other hand, the capacities of methanation and direct air capture plants increase by $20$\% and $11$\%, respectively, which slightly offsets the cost savings made elsewhere. Although not shown in Figure \ref{sensitivitycapacities}, the capacity of liquefied methane storage tanks in the coastal hub more than doubles. This can be explained by the fact that the transport of liquefied methane does not occur on a continuous basis, which introduces some inflexibility in the supply chain. Since liquefied methane storage is much cheaper than battery or hydrogen storage, the buffer absorbing the variability of renewable power generation is moved downstream in the supply chain. It is worth noting that even if the transport of liquefied methane took place continuously, the mismatch between the production and demand profiles would need to be balanced by some storage capacity, which will always come at a cost.

\textbf{Investment Costs} A number of technologies used in the proposed remote carbon-neutral synthetic methane supply chain have not yet reached full maturity. In particular, electrolysis (EL), direct air capture and methanation plants are still undergoing development, and their costs thus remain highly uncertain to 2030. The cost figures used for electrolysis plants in the reference scenario lean towards the moderately optimistic side for 2030, and achieving them would require both sustained research and development efforts and a commercial scale-up \citep{Schmidt2017}. Likewise, the cost figures used for direct air capture plants are on the fairly optimistic side for 2030. Indeed, the cost of each ton of carbon dioxide captured from the atmosphere is between 65 and 70 \euro\mbox{ } (energy costs excluded) in the reference scenario, which is close to long-term cost targets of \$100/ton (energy costs included) that \cite{Keith2018} seek to reach. Methanation cost estimates, however, are rather conservative \citep{methanationcosts,solarpvcosts}. The uncertainty around these costs and their impact on synthetic methane cost is resolved as follows. Firstly, the CAPEX and FOM of electrolysis and direct air capture plants are increased by 50\%, which yields synthetic methane cost estimates around 168.1 \euro/MWh (which is approximately $12$\% higher than that of the reference scenario).  Then, the CAPEX and FOM of electrolysis, direct air capture and methanation plants are decreased by 50\%, first on an individual basis and then all at once. Although such drastic cost reductions seem very unlikely, they nevertheless make it possible to estimate the sensitivity of synthetic methane cost to these assumptions and provide a lower bound on costs that may realistically be achieved. The least sensitive of these parameters is the cost of direct air capture plants, followed by the cost of methanation plants. Decreasing them individually only leads to methane cost reductions of $5$\% or less. Decreasing the cost of electrolysis plants by $50$\%, however, has a much greater impact and leads to cost savings around $10$\%. Decreasing the costs of all three technologies at once leads to cost savings of roughly $16$\% and yields a strong lower bound of $125.1$\euro/MWh on the reductions in synthetic methane cost that may be achieved in this fashion. It is worth noting that these cost savings are achieved with virtually no change in deployed capacities (as shown in Figure \ref{sensitivitycapacities}), which suggests that the latter mostly depend on renewable production profiles, the flexibility of technologies and the demand that must be satisfied.

\textbf{DAC Energy Consumption} The direct air capture process used in this analysis requires high-temperature heat to calcine calcium carbonate compounds and release the carbon dioxide that they trap. In this paper, it is assumed that this heat is provided by burning hydrogen, and slightly less than 20\% of the total hydrogen production is used to this end. Hence, the fact that this hydrogen must be produced from renewable electricity leads to the deployment of additional power generation, storage, transport, electrolysis and hydrogen storage capacity, which directly translates into a lower overall efficiency of the full supply chain and higher synthetic methane costs. Instead, a different capture process that only uses low to medium-temperature heat and electricity could be used \citep{Wurzbacher2011}. It has been estimated that this process consumes approximately $0.5$ MWh of electricity per ton of carbon dioxide captured and requires roughly $2.5$ MWh of heat at 100\degree C. Since the production of one ton of synthetic methane by the Sabatier reaction releases $2.87$ MWh of high temperature heat (and requires $2.75$ tons of carbon dioxide), some of the heat required by the direct air capture process could be supplied by nearby methanation plants. The impact of such a change is analysed by increasing the electricity consumption of the direct air capture process fivefold (i.e., setting it to roughly $0.5$ MWh/ton instead of $0.1$ MWh/ton in the reference scenario) and setting the hydrogen consumption to zero. The same cost figures those of the reference scenario are used. Results shown in Figures \ref{sensitivitycosts} and \ref{sensitivitycapacities} confirm the intuition that using hydrogen to satisfy the heat requirements of a high-temperature direct air capture process has a substantial impact on synthetic methane cost and the capacities of various key technologies. More precisely, the cost of synthetic methane in this configuration is approximately 10\% lower than that found in the reference scenario, while the capacities of power generation and electrolysis plants are $10$\% and $20$\% smaller, respectively. Although promising, a detailed analysis of the heat integration potential and the cost of this process should be performed in order to confirm these findings.

\textbf{Financing Costs} A weighted average cost of capital of $7$\% has so far been used. In order to evaluate the impact of financing costs, a hypothetical situation where the cost of financing the system is set to zero is now studied (i.e., annualised CAPEX values are calculated using Eq. \ref{nowaccformula}). Thus, in this set-up, the cost of synthetic methane production and delivery solely reflects the cost and efficiency of technologies in the supply chain, and provides an absolute lower bound on costs that may be realistically achieved. Results in Figure \ref{sensitivitycosts} suggest that neglecting financing costs leads to a synthetic methane cost of $88.3$ \euro/MWh (which corresponds to a $40$\% reduction compared with the reference scenario), which is by far the lowest observed in this paper and highlights the influence of weighted average cost of capital assumptions. It is also worth noting that this cost decrease is achieved with little change in the capacities of conversion and storage technologies compared with the reference scenario.

\subsection{Discussion}\label{discussion}

Discrepancies exist between the results presented in Sections \ref{results_ref_scen} and \ref{results_sensitivity} and synthetic methane production cost estimates published elsewhere in the literature. Indeed, recall that \cite{Zeman2008} provide cost estimates ranging from 74.1 to 94.6  \euro/MWh. Furthermore, in \cite{Fasihi2015}, the cost of producing synthetic methane from renewable electricity in central and southern Algeria and delivering it to Japan is estimated to be around 65-75 \euro/MWh in 2030 for a hybrid solar-wind system using a WACC of 7\%. In \cite{Fasihi2017}, the cost of producing synthetic methane in the same region and delivering it to Finland is estimated to be between 100-110 \euro/MWh (HHV) by 2030 and between 90-100 \euro/MWh (HHV) by 2040, respectively, using a WACC of 7\%. Finally, in \cite{solarpvcosts}, a uniform WACC of 6\% is used, yielding cost estimates around 140 \euro/MWh (LHV) for a solar PV configuration and around 150 \euro/MWh (LHV) for a hybrid solar-wind configuration. It is worth noting that the hybrid solar-wind configuration is slightly more expensive than the solar-powered system in their reference cost scenario.

The methods used in the aforementioned papers, which are discussed in Section \ref{related_work}, suffer from several shortcomings. More precisely, they use a very low temporal resolution (one time period per year in the so-called full load hour model) that completely smoothes out the variability of power production signals. Furthermore, their models have a very low level of technical detail. The combination of these two features makes it very difficult to capture the interaction between subsystems and accurately model the supply chain in an integrated fashion. Hence, this typically removes the need for oversizing renewable power generation technologies or deploying flexibility options such as storage systems to balance the variable power supply and satisfy operating constraints, in spite of the fact that the operation of some technologies further down the chain is inflexible or discontinuous (e.g., methanation plants or transport by liquefied methane carrier vessels). Oversizing plants or deploying storage technologies is relatively expensive and both account for a non-negligible share of the final methane cost, as discussed in Sections \ref{results_ref_scen} and \ref{results_sensitivity}. For example, the fact that solar PV variability has been completely smoothed out by the full load hour model used in \cite{solarpvcosts} explains the fact that solar-only and hybrid solar-wind configurations yield very close methane cost estimates, while the solar-only configuration is almost 35\% more expensive than the hybrid wind-solar configuration considered in this paper. Thus, the aforementioned papers underestimate final product cost as a result of inadequate modelling choices. In addition, some of the techno-economic assumptions made in \cite{Fasihi2015,Fasihi2017} seem particularly optimistic. For example, the CAPEX of electrolysis and methanation plants is approximately two and three times lower than the values used in the reference case presented in this paper, respectively. These assumptions clearly lead to low methane cost estimates but are poorly supported. Indeed, to the authors' best knowledge, such assumptions do not appear elsewhere in the literature or in publicly-accessible databases and are therefore difficult to cross-check.

\section{Conclusion and Future Work}
\label{conclusion}

This paper studies the economics of carbon-neutral synthetic fuel production in remote areas where high-quality renewable resources are abundant. With this goal in mind, a (hyper)graph-based optimisation modelling framework directly applicable to the strategic planning of remote renewable energy supply chains is proposed. The method is leveraged to study the economics of carbon-neutral synthetic methane production from solar and wind energy in North Africa. 

The full supply chain is modelled and optimised in an integrated fashion over five years (2015-2019) with hourly time resolution. Essential operational constraints are taken into account, which is key for accurately capturing interactions between subsystems. Results suggest that the cost of synthetic methane delivery to northwestern European consumers would be around $149.7$ \euro/MWh (HHV) by 2030 for a system that relies on a combination of solar photovoltaic and wind power plants, assuming a uniform weighted average cost of capital of 7\%. A comprehensive sensitivity analysis has also been carried out in order to evaluate the impact of various techno-economic parameters and assumptions on synthetic methane cost, including the availability of wind power plants, the investment costs of electrolysis, methanation and direct air capture plants, their operational flexibility, the energy consumption of direct air capture plants, and financing costs. The most expensive configuration (around 200 \euro/MWh) relies on solar photovoltaic power plants alone, while the cheapest configuration (around 88 \euro/MWh) makes use of a combination of solar PV and wind power plants and is obtained when financing costs are set to zero. The cost estimates found for the reference scenario and the configuration relying solely on solar PV power plants are much higher than those previously published in the literature. This discrepancy can be partly explained by the fact that the models used in previous studies had a very low temporal resolution and failed to properly capture the interactions between highly variable power generation plants (especially solar photovoltaic units) and inflexible conversion technologies (such as methanation plants) and demand profiles.

Several research directions can be pursued in future work. Firstly, quantitatively analysing some of the options suggested for cost reductions would provide more insight into the economic potential of an energy supply pathway based on carbon-neutral methane synthesis in remote areas. Then, leveraging the framework to study different pathways involving different regions (and thus resource types and profiles) and energy carriers (e.g., hydrogen, methanol or ammonia), would allow one to draw a complete picture of energy supply options and to compare their respective merits. Finally, the graph-based modelling framework could be expanded in different ways. For instance, the class of problems that can be represented could be broadened by introducing nonlinear expressions. The graph representation could also be exploited to facilitate preprocessing tasks and the analysis of model properties, eventually opening the door to the deployment of more efficient solution methods that better exploit problem structure \citep{Jalving2019}.

\newpage
\section{Nomenclature}
\begin{table*}[h!]
\begin{tabular}{|c|c|}
		\hline
		 & \textbf{\small{Sets and Indices}}\\
		\hline
		\small{$\mathcal{E}, e$} & set of hyperedges and hyperedge index\\
		\small{$e_T, e_H$} & tail and head of hyperedge $e \in \mathcal{E}$\\
		\small{$\mathcal{G}$} & hypergraph with node set $\mathcal{N}$ and hyperedge set $\mathcal{E}$\\
		\small{$\mathcal{I}^n, i$} & set of external variables at node $n$, and variable index\\
		\small{$\mathcal{N}, n$} & set of nodes and node index\\
		\small{$\mathcal{T}, t$} & set of time periods and time index\\
		\hline
		& \textbf{\small{Parameters}}\\
		\hline
		\small{$\nu \in \mathbb{N}$} & number of years spanned by optimisation horizon \\
		\small{$\pi_{t}^n \in [0,1]$} & (operational) availability of conversion node $n$ at time $t$ \\
		\small{$\underbar{$\kappa$}^{n} \in \mathbb{R}_+$} & existing flow capacity of conversion or storage node $n$\\
		\small{$\bar{\kappa}^n \in \mathbb{R}_+$} & maximum flow capacity of conversion or storage node $n$\\
		\small{$\mu^{n} \in [0,1]$} & minimum operating level of conversion node $n$ (fraction of capacity)\\
		\small{$\delta t \in \mathbb{R}_+$} & duration of each time period\\
		\small{$\Delta_{i,+}^n \in [0,1]$} & maximum ramp-up rate for flow $i$ and conversion node $n$ (frac. of capacity per unit time)\\
		\small{$\Delta_{i,-}^n \in [0,1]$} & maximum ramp-down rate for flow $i$ and conversion node $n$ (frac. of capacity per unit time)\\
		\small{$\phi_{i}^n \in \mathbb{R}_+$} & conversion factor between reference flow $r$ and flow $i$ for conversion or storage node $n$\\
		\small{$\tau_{i}^n \in \mathbb{N}$} & conversion time delay for flow $i$ of conversion node $n$\\
		\small{$\eta_{S}^n \in [0, 1]$} & self-discharge rate of storage node $n$\\
		\small{$\eta_{+}^n \in [0, 1]$} & charge efficiency of storage node $n$\\
		\small{$\eta_{-}^n \in [0, 1]$} & discharge efficiency of storage node $n$\\
		\small{$\sigma^{n} \in [0, 1]$} & minimum inventory level of storage node $n$ (fraction of capacity)\\
		\small{$\bar{\epsilon}^{n} \in \mathbb{R}_+$} & maximum inventory capacity of storage node $n$\\
		\small{$\underbar{$\epsilon$}^{n} \in \mathbb{R}_+$} & existing inventory capacity of storage node $n$\\
		\small{$\rho^{n} \in \mathbb{R}_+$} & maximum discharge-to-charge ratio of storage node $n$\\
		\small{$\lambda_{t}^e \in \mathbb{R}$} & withdrawal/injection at time $t$ and conservation hyperedge $e$\\
		\small{$\zeta^{n} \in \mathbb{R}_+$} & annualised CAPEX of node $n$ (flow component)\\
		\small{$\theta_{f}^n \in \mathbb{R}_+$} & FOM cost of of node $n$ (flow component)\\
		\small{$\theta_{t,v}^n \in \mathbb{R}_+$} & VOM cost of node $n$ (flow component)\\
		\small{$\varsigma^{n} \in \mathbb{R}_+$} & annualised CAPEX of storage node $n$ (stock component)\\
		\small{$\vartheta_{f}^n \in \mathbb{R}_+$} & FOM cost of storage node $n$ (stock component) \\
		\small{$\vartheta_{t,v}^n \in \mathbb{R}_+ $} & VOM cost of storage node $n$ (stock component)\\
		\small{$\theta_{t,L}^n \in \mathbb{R}_+$} & cost of unserved demand at conservation node $n$\\
		\hline
		 & \textbf{\small{Variables}}\\
		\hline 
		\small{$q_{it}^n \in \mathbb{R}_+$} & flow variable $i$ of node $n$ at time $t$\\
		\small{$K^{n} \in \mathbb{R}_+$} & new flow capacity of node $n$\\
		\small{$e_{t}^n \in \mathbb{R}_+$} & inventory level of storage node $n$ at time $t$\\
		\small{$E^{n} \in \mathbb{R}_+$} & new stock capacity of storage node $n$\\
		\hline
	\end{tabular}
\end{table*}

\section*{Conflict of Interest Statement}
%All financial, commercial or other relationships that might be perceived by the academic community as representing a potential conflict of interest must be disclosed. If no such relationship exists, authors will be asked to confirm the following statement: 

The authors declare that the research was conducted in the absence of any commercial or financial relationships that could be construed as a potential conflict of interest.

\section*{Author Contributions}

Mathias Berger and Damien Ernst designed the research. Mathias Berger, David Radu and Ghislain Detienne collected the data. Mathias Berger performed the research and drafted the manuscript. David Radu, Ghislain Detienne, Thierry Deschuyteneer, Aurore Richel and Damien Ernst provided feedback on the research and manuscript.

%\section*{Funding}
%Details of all funding sources should be provided, including grant numbers if applicable. Please ensure to add all necessary funding information, as after publication this is no longer possible.

\section*{Acknowledgments}
The authors would like to thank Adrien Bolland, Hatim Djelassi and Virginie Pison for providing feedback on an earlier version of this manuscript. The authors would also like to thank Julien Confetti for his precious help with the design of figures and diagrams in this paper. Finally, the authors would like to gratefully acknowledge the support of the Federal Government of Belgium through its Energy Transition Fund and the INTEGRATION project.

\section*{Data Availability Statement}
The datasets and code used for this study can be found in the GBOML repository \citep{GBOML}.

\bibliographystyle{frontiersinSCNS_ENG_HUMS} 
\bibliography{frontiers_bib}

\end{document}